\documentclass[prb,twocolumn,superscriptaddress]{revtex4-1}
\usepackage[usenames,dvipsnames]{color}
\usepackage{amsmath}
\usepackage{amssymb}
\usepackage{graphicx}
\usepackage{tabularx}
\usepackage{color}
\usepackage{ulem}
\usepackage[bookmarks=true,colorlinks=true,linkcolor=blue,urlcolor=blue,
citecolor=blue]{hyperref}
\usepackage{epstopdf}
\begin{document}

\title{Investigation of the effective interactions for the Emery model by the
constrained random-phase approximation and constrained functional
renormalization group}
\author{Xing-Jie Han}

\affiliation{Institute for Theoretical Solid State Physics, RWTH Aachen
University, D-52056 Aachen, Germany}

\author{Philipp Werner}

\affiliation{Department of Physics, University of Fribourg, 1700 Fribourg, Switzerland}

\author{Carsten Honerkamp}

\affiliation{Institute for Theoretical Solid State Physics, RWTH Aachen
University, and JARA Fundamentals of Future Information Technology, D-52056
Aachen, Germany}

\date{\today }

\begin{abstract}
The effective interaction of downfolded low-energy models  for electrons in solids can be obtained by integrating out the high energy bands away from the target band near the
Fermi level. Here, we apply the constrained random-phase
approximation (cRPA) and constrained functional renormalization group
(cfRG), which can go beyond cRPA by including all one-loop diagrams, to
calculate and compare the effective interactions of the three-band Emery
model, which is often used to investigate cuprate high-temperature superconductors. At half band filling, we find that the effective interaction increases as
the charge transfer energy ($\Delta _{dp}$) increases and similar behavior is
obtained as a function of the interatomic 2$p$-3$d$ interaction ($U_{dp}$).
However, the effective interaction is more sensitive to $\Delta _{dp}$ than $%
U_{dp}$. For most of the parameter sets, the effective static interaction is
overscreened in cRPA compared to cfRG. The low-energy models at half-filling are solved within dynamical mean-field theory (DMFT). The results show that despite the different static interactions, the systems with cRPA and cfRG interaction exhibit a Mott transition at similar values of $\Delta _{dp}$. We also investigate the effective interaction as
a function of doping. The cfRG effective interaction decreases as the electron
number increases and displays a trend opposite to that of cRPA.
Antiscreening is observed for the hole-doped case. For all the cases
studied, the near-cancellation of the direct particle-hole channel is
observed. This indicates that at least for the downfolding of the onsite interaction terms, methods beyond cRPA may be 
required.
\end{abstract}

\maketitle

\preprint{Draft}

\section{Introduction}

Tremendous progress has been made in understanding and predicting new
materials with particular properties by \textit{ab initio} methods based on
density functional theory (DFT),\cite{Kohn-1965} including recent advances in
topological semimetals\cite{Wan-2011,Armitage-2018} and superconductivity in
the sulfur hydride systems.\cite{Li-2014,Drozdov-2015} Despite the success
of DFT for those weakly correlated materials, the method suffers from intrinsic
difficulties and sometimes gives qualitatively wrong results for strongly
correlated materials. For example, the insulating behavior of some
transition metal oxides is not captured by DFT.\cite%
{Pickett-1989,Imada-1998} In strongly correlated systems, the Coulomb repulsion
between the low-energy electrons occupying the partially filled narrow bands
in proximity to the Fermi level is comparable or larger than the bandwidth,
and in such a situation the independent-particle picture of DFT breaks down. Meanwhile,
most macroscopic properties of materials are mainly determined by those
low-energy states near the Fermi level at low temperatures. Therefore, to
accurately account for the low-energy correlation effects beyond the
mean-field level, more sophisticated numerical methods aimed at the study of
many-body model Hamiltonians, like the fluctuation exchange approximation (FLEX),%
\cite{Savrasov-2018,Sakakibara-2019} Quantum Monte Carlo (QMC) methods\cite%
{Tahara-2008,Imada-2010,Ma-2015} and Dynamical Mean Field Theory (DMFT),\cite%
{Metzner-1989,Georges-1992,Georges-1996} have been employed to complement DFT
calculations. The combination of DFT and DMFT has been extensively used to
study correlated materials ranging from high-temperature superconductors to
heavy fermion systems.\cite{Kotliar-2006,Goremychkin-2018,Adler-2019}

A crucial step in most of the DFT+DMFT calculations is the construction of
low-energy effective Hamiltonians in the subspace of relevant states near the Fermi level,
which contains the correlated $d$ or $f$ orbitals, starting from the complete
Hilbert space. In this so-called downfolding process, the parameters of
the low-energy effective Hamiltonians are derived by integrating out the
high-energy degrees of freedom successively, in the spirit of renormalization
group methods.\cite{Wilson-1975} While the downfolding concept is physically
intuitive and straightforward, the parameter-free determination of the
(partially) screened low-energy interactions still remains a major challenge
and bottleneck for practical \textit{ab initio} calculations.

To compute the screened Coulomb interaction for the low-energy effective
Hamiltonians, a commonly used method is the constrained-random-phase
approximation (cRPA).\cite{Aryasetiawan-2004,Miyake-2009} In cRPA, the
partially screened Coulomb interaction between electrons in the low-energy
subspace is frequency-dependent and is determined by $U\left( \omega
\right) =\left[ 1-\upsilon P_{r}\left( \omega \right) \right] ^{-1}\upsilon $%
, where $\upsilon$ is the bare Coulomb interaction and $P_{r}$ is the
particle-hole polarization function calculated by the RPA-type diagrams,
excluding the contributions solely within the low-energy subspace.
The latter is needed because polarization effects within the low-energy subspace are
explicitly treated in the solution of the low-energy effective Hamiltonian. More
specifically, at least one of the lines in the polarization particle-hole
bubble $P_{r}$ belongs to the high-energy sector. Although cRPA appears to
be a big step forward and has been applied to calculate the screened Coulomb
interactions for various materials and toy models,\cite%
{Aryasetiawan-2006,Nakamura-2009,Wehling-2011,Friedrich-2011,Vaugier-2012,Casula-2012,Shinaoka-2015,Han-2018,Hirayama-2020}
both underestimations\cite{Nakamura-2009} and overestimations\cite%
{Casula-2012,Shinaoka-2015,Han-2018} of the screening effects have been reported.
Since many physical properties of strongly correlated materials are
sensitive to small changes in $U$, especially near phase transition points, a
better understanding of the limitations of cRPA is important for reliable
parameter-free \textit{ab initio} calculations and predictions.

Recently, the constrained functional renormalization group (cfRG) method\cite%
{Honerkamp-2012,Kinza-2015,Honerkamp-20181,Honerkamp-2018} has been proposed
to go beyond cRPA by including all one-loop diagrams and vertex corrections.
The cRPA can be recovered by only keeping the RPA-type diagrams. A
multi-orbital Hubbard model which could be solved exactly using QMC has been
studied to test the accuracy of cRPA and cfRG at the level of the effective
one-band interaction after downfolding.\cite{Honerkamp-2018} Significant
corrections to cRPA were found and the overestimated screening effects in
cRPA could be explained by the near-cancellation of certain loop corrections
in cfRG. Furthermore, antiscreening effects, i.~e. an enhancement of
the bare interactions, were predicted by cfRG for the cases studied in Refs.~\onlinecite{Kinza-2015,Honerkamp-20181,Honerkamp-2018},  in contrast to the
general suppression of the bare interactions by cRPA.

In this work, to further explore the differences between cRPA and
cfRG, we use cRPA and cfRG downfolding schemes for the two-dimensional
three-band Emery model\cite{Emery-1987} and calculate the effective
interactions for the low-energy band using different model parameter sets.
To further reveal the origin of the different results obtained by these two
methods, the idea of channel decomposition is used.
The effects of the resulting frequency-dependent cfRG and cRPA interactions
are illustrated by solving the corresponding low-energy single band models
using DMFT.

The rest of the paper is organized as follows. In Sec.~\ref{method} we
introduce the Emery model and present the cRPA and cfRG downfolding schemes.
Section~\ref{results} presents the cRPA and cfRG calculations for the effective interactions
in the target band. The DMFT results are discussed in Sec. \ref{DMFT}, and the paper is
summarized in Sec.~\ref{conclusion}.

\section{MODEL AND METHODS}

\label{method}

\subsection{Model Hamiltonian}

\begin{figure}[tbp]
\includegraphics[width=0.99\columnwidth]{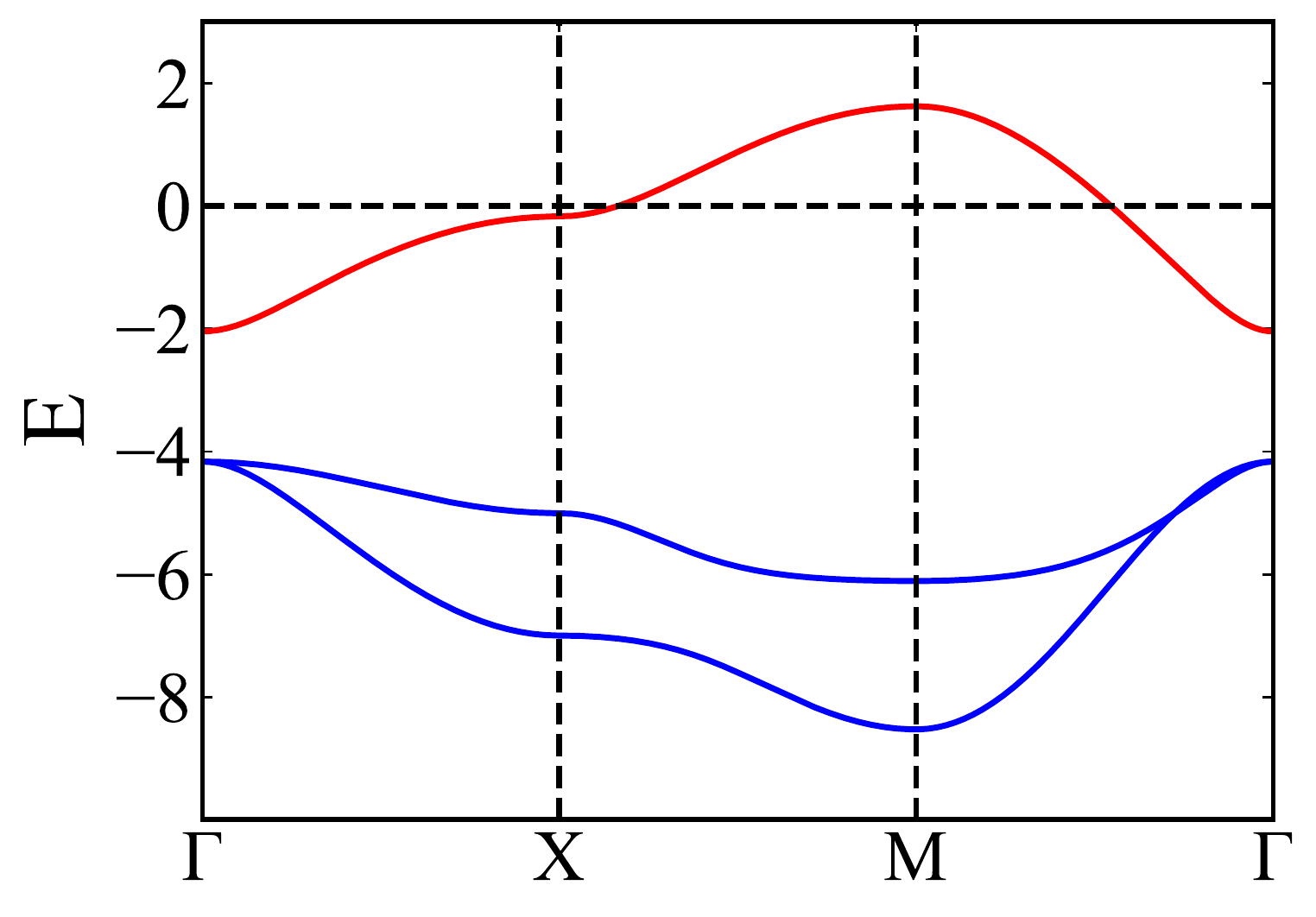}
\caption{Electronic band structure of the half-filled Emery model along the
high-symmetry directions $\Gamma \rightarrow X\rightarrow M\rightarrow
\Gamma $ in the Brillouin zone of the square lattice with the 1MTO hopping
parameters and $\Delta _{dp}=2$. The high-symmetry points are defined as $%
\Gamma =\left( 0,0\right) $, $X=\left( \protect\pi ,0\right) $ and $M=\left(
\protect\pi ,\protect\pi \right) $. Energy zero corresponds to the Fermi
level. The red line represents the target band and the two blue lines show
the screening bands.}
\label{Disper}
\end{figure}

Compared to the extensively investigated one-band Hubbard model, a more
realistic model to describe the physics of cuprates is the three-band Emery
model, which includes the copper 3$d_{x^{2}-y^{2}}$ orbital and the ligand
oxygen 2$p_{x}$ and 2$p_{y}$ orbitals for the CuO$_{2}$ planes. It is
defined by the Hamiltonian

\begin{eqnarray}
H &=&\sum_{\mathbf{k},\sigma }c_{\mathbf{k},\sigma }^{\dag }h_{0}\left(
\mathbf{k}\right) c_{\mathbf{k},\sigma }  \notag \\
&&+U_{dd}\sum_{i}n_{i,\uparrow }^{d}n_{i,\downarrow
}^{d}+U_{pp}\sum_{j}n_{j,\uparrow }^{p_{j}}n^{p_{j}}_{j,\downarrow }  \notag \\
&&+U_{dp}\sum_{\left\langle i,j\right\rangle ,\sigma ,\sigma ^{\prime
}}n_{i\sigma }^{d}n_{j\sigma ^{\prime }}^{p_{j}} \,, \label{Hamiltonian}
\end{eqnarray}%
where $c_{\mathbf{k},\sigma }^{\dag }=(d_{\mathbf{k},\sigma }^{\dag },p_{x,%
\mathbf{k},\sigma }^{\dag },p_{y,\mathbf{k},\sigma }^{\dag })$ is the
creation operator for the electrons on 3$d_{x^{2}-y^{2}}$, 2$p_{x}$ and 2$%
p_{y}$ orbitals. $h_{0}\left( \mathbf{k}\right) $ is the kernel of the
noninteracting tight-binding Hamiltonian in momentum space given in Ref.~\onlinecite{Hansmann-2014}.
The hopping parameters are chosen to be those
of the 1MTO model. The diagonal elements of $h_{0}\left( \mathbf{k}\right) $
contain the information about the charge-transfer energy, which is defined
as the energy separation between the copper 3$d_{x^{2}-y^{2}}$ orbital and
the oxygen 2$p$ orbital, $\Delta _{dp}=\varepsilon _{d}-\varepsilon _{p}$. In
this study, we report all energies in eV.

The half-filled noninteracting band structure for $\Delta _{dp}=2$ is
plotted in Fig.~\ref{Disper} along some specific high-symmetry directions in
the Brillouin zone of the square lattice. The half-filled upper band (red
line) which crosses the Fermi level is our low-energy target band and the
remaining two fully filled bands (blue lines) are the high-energy bands to be
integrated out successively. As can be seen, for the Emery model, the
high-energy and low-energy degrees of freedom are well separated by $\Delta
_{dp}$, making the procedure to integrate out the high-energy bands
well-defined. At the same time, $\Delta _{dp}$ is roughly the magnitude of the
minimum gap between the target band and the screening bands, which is the
main parameter in the polarization functions and plays an important role
during the renormalization process.\cite{Honerkamp-2012}

The interaction term of the Hamiltonian in Eq.~(\ref{Hamiltonian}) is given
in real space. Here, the labels $i$ and $j$ run over all Cu and O sites, and $%
\left\langle i,j\right\rangle $ denotes the summation over nearest neighbor
Cu-O bonds. The interaction parameters $U_{dd}$ and $U_{pp}$ correspond to
the on-site Coulomb repulsion between two electrons with opposite spins
located on the Cu 3$d$ and O 2$p$ orbitals, respectively. The nearest
neighbor Cu-O interaction $U_{dp}$ is neglected in many studies but has been shown to
be important to stabilize the charge transfer insulating state of the Emery
model at half filling.\cite{Hansmann-2014,Werner-2015} In the following
study, we use $U_{dd}=13$ and $U_{pp}=7$. The values of $\Delta _{dp}\ $and $%
U_{dp}$ will be varied and specified explicitly in the following subsections.

\subsection{cfRG procedure}

In this paper, we only study the SU(2)-symmetric case and write the
two-particle interaction as $V_{o_{1}o_{2};o_{3}o_{4}}^{\Lambda }\left(
k_{1},k_{2},k_{3},k_{4}\right) $, where $\Lambda $ is the flow parameter.
The variables $o_{i}$ denote orbital indices and $k_{i}=(\mathbf{k}_{i},i\omega
_{i})$ combined momenta and Matsubara frequencies. The indices 1 and 2
denote the two incoming particles while 3 and 4 denote the two outgoing
particles. Here, the SU(2)-symmetry is fulfilled by the requirement that
the configuration for the spin indices of the interaction is fixed by the
Kronecker delta $\delta _{s_{1},s_{3}}\delta _{s_{2},s_{4}}$, which means
the incoming particle 1 and the outgoing particle 3 have the same spin
projection and particles 2 and 4 have the same spin projection.\cite%
{Salmhofer-2001} Since $k_{4}$ can be determined by momentum/energy
conservation, $k_{4}=k_{1}+k_{2}-k_{3}$, the two-particle vertex is denoted
as $V_{o_{1}o_{2};o_{3}o_{4}}^{\Lambda }\left( k_{1},k_{2},k_{3}\right) $ in
the following to abbreviate the notation.

\begin{figure}[tbp]
\includegraphics[width=0.99\columnwidth]{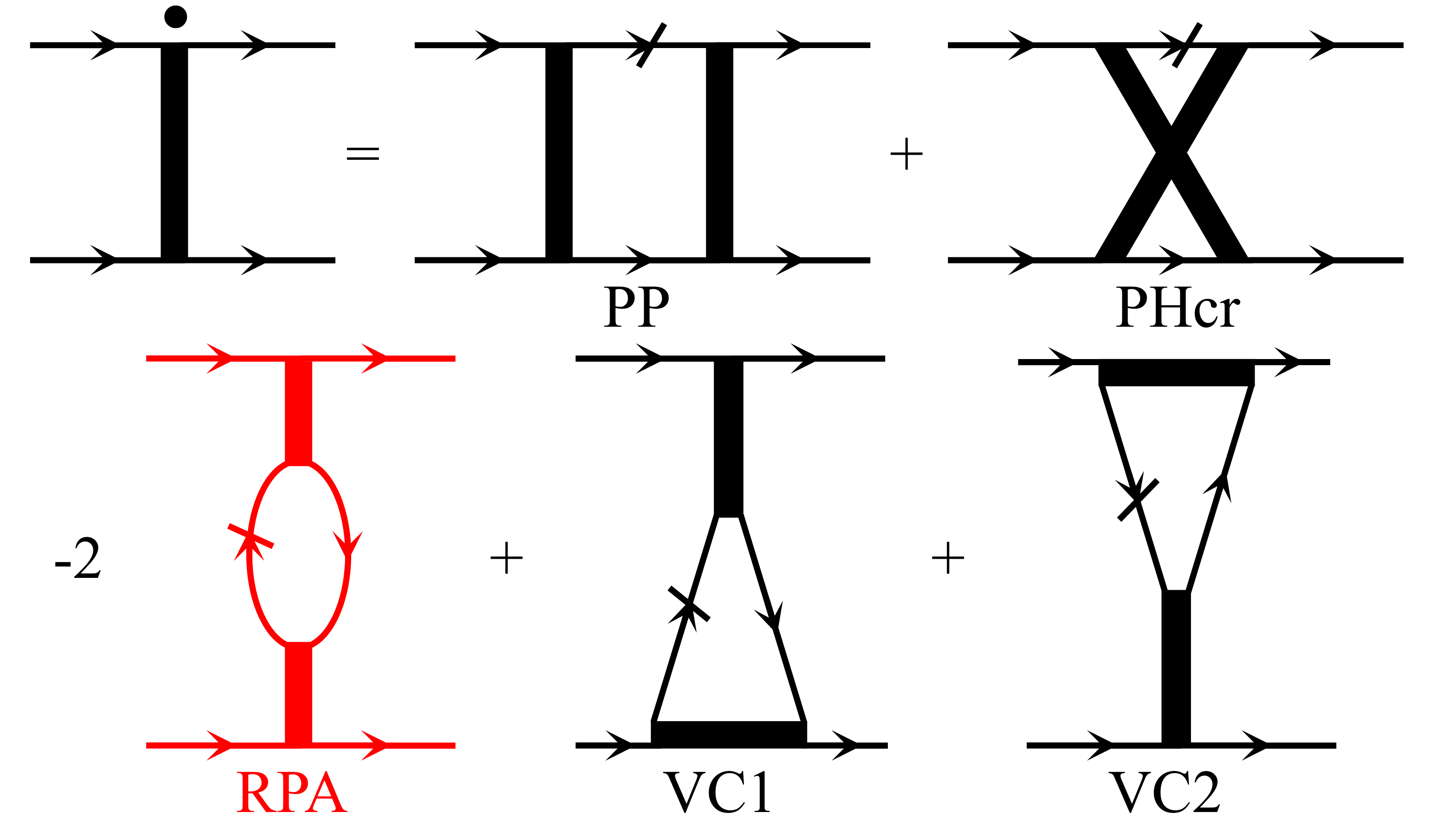}
\caption{Diagrammatic representation of the fRG flow equations. The cRPA is
equivalent to the red diagram denoted as RPA. The dot over the interaction
vertex on the left-hand side of the flow equation denotes the derivative
with respect to the cutoff $\Lambda $. The internal solid line represents the
free propagator and the slashed solid line denotes the corresponding
single-scale propagator.}
\label{Feynman}
\end{figure}

Without self-energy corrections, the SU(2) symmetric flow equations in the
level-2 truncation for the two-particle interaction in the cfRG scheme can
be obtained by employing the Wick-ordered fRG formalism\cite{Honerkamp-2012}
and read as follows:

\begin{eqnarray}
&&\partial _{\Lambda }V_{o_{1}o_{2};o_{3}o_{4}}^{\Lambda }\left(
k_{1},k_{2},k_{3}\right) =\partial _{\Lambda
}P_{o_{1}o_{2};o_{3}o_{4}}^{\Lambda }\left( k_{1},k_{3};s\right)  \notag \\
&&\hspace{10mm}+\partial _{\Lambda }D_{o_{1}o_{2};o_{3}o_{4}}^{\Lambda }\left(
k_{1},k_{4};t\right) +\partial _{\Lambda }C_{o_{1}o_{2};o_{3}o_{4}}^{\Lambda
}\left( k_{1},k_{3};u\right) ,\nonumber\\  \label{cfrg_flow}
\end{eqnarray}%
where the Mandelstam variables are introduced as

\begin{equation}
s=k_{1}+k_{2}, \, t=k_{3}-k_{1}, \, u=k_{4}-k_{1}.
\end{equation}%
The three terms on the right-hand side of the flow equation Eq.~(\ref%
{cfrg_flow}) are called the particle-particle channel (P), the direct
particle-hole channel (D) and the crossed particle-hole channel (C),
respectively. The expressions are given by (using $N$ for the number of unit cells)%
\begin{widetext}
\begin{eqnarray}
\partial _{\Lambda }P_{o_{1}o_{2};o_{3}o_{4}}^{\Lambda }\left(
k_{1},k_{3};s\right)  &=&\frac{T}{N}\sum_{\substack{ k \\ \tilde{o}_{1}%
\tilde{o}_{2};\tilde{o}_{3}\tilde{o}_{4}}}V_{o_{1}o_{2};\tilde{o}_{1}\tilde{o%
}_{2}}^{\Lambda }\left( k_{1},s-k_{1},k\right) \partial _{\Lambda }\left[ G_{%
\tilde{o}_{1}\tilde{o}_{3}}^{\Lambda }\left( k\right) G_{\tilde{o}_{2}\tilde{%
o}_{4}}^{\Lambda }\left( s-k\right) \right] V_{\tilde{o}_{3}\tilde{o}%
_{4};o_{3}o_{4}}^{\Lambda }\left( k,s-k,k_{3}\right) ,  \label{PP} \\
\partial _{\Lambda }D_{o_{1}o_{2};o_{3}o_{4}}^{\Lambda }\left(
k_{1},k_{4};t\right)  &=&-\frac{2T}{N}\sum_{\substack{ k \\ \tilde{o}_{1}%
\tilde{o}_{2};\tilde{o}_{3}\tilde{o}_{4}}}V_{o_{1}\tilde{o}_{4};o_{3}\tilde{o%
}_{1}}^{\Lambda }\left( k_{1},k+t,k_{1}+t\right) \partial _{\Lambda }\left[
G_{\tilde{o}_{1}\tilde{o}_{3}}^{\Lambda }\left( k\right) G_{\tilde{o}_{2}%
\tilde{o}_{4}}^{\Lambda }\left( k+t\right) \right] V_{\tilde{o}_{3}o_{2};%
\tilde{o}_{2}o_{4}}^{\Lambda }\left( k,k_{4}+t,k+t\right)   \notag \\
&&+\frac{T}{N}\sum_{\substack{ k \\ \tilde{o}_{1}\tilde{o}_{2};\tilde{o}_{3}%
\tilde{o}_{4}}}V_{o_{1}\tilde{o}_{4};o_{3}\tilde{o}_{1}}^{\Lambda }\left(
k_{1},k+t,k_{1}+t\right) \partial _{\Lambda }\left[ G_{\tilde{o}_{1}\tilde{o}%
_{3}}^{\Lambda }\left( k\right) G_{\tilde{o}_{2}\tilde{o}_{4}}^{\Lambda
}\left( k+t\right) \right] V_{\tilde{o}_{3}o_{2};o_{4}\tilde{o}%
_{2}}^{\Lambda }\left( k,k_{4}+t,k_{4}\right)   \notag \\
&&+\frac{T}{N}\sum_{\substack{ k \\ \tilde{o}_{1}\tilde{o}_{2};\tilde{o}_{3}%
\tilde{o}_{4}}}V_{o_{1}\tilde{o}_{4};\tilde{o}_{1}o_{3}}^{\Lambda }\left(
k_{1},k+t,k\right) \partial _{\Lambda }\left[ G_{\tilde{o}_{1}\tilde{o}%
_{3}}^{\Lambda }\left( k\right) G_{\tilde{o}_{2}\tilde{o}_{4}}^{\Lambda
}\left( k+t\right) \right] V_{\tilde{o}_{3}o_{2};\tilde{o}%
_{2}o_{4}}^{\Lambda }\left( k,k_{4}+t,k+t\right) ,  \label{DPH} \\
\partial _{\Lambda }C_{o_{1}o_{2};o_{3}o_{4}}^{\Lambda }\left(
k_{1},k_{3};u\right)  &=&\frac{T}{N}\sum_{\substack{ k \\ \tilde{o}_{1}%
\tilde{o}_{2};\tilde{o}_{3}\tilde{o}_{4}}}V_{o_{1}\tilde{o}_{4};\tilde{o}%
_{1}o_{4}}^{\Lambda }\left( k_{1},k+u,k\right) \partial _{\Lambda }\left[ G_{%
\tilde{o}_{1}\tilde{o}_{3}}^{\Lambda }\left( k\right) G_{\tilde{o}_{2}\tilde{%
o}_{4}}^{\Lambda }\left( k+u\right) \right] V_{\tilde{o}_{3}o_{2};o_{3}%
\tilde{o}_{2}}^{\Lambda }\left( k,k_{3}+u,k_{3}\right) .  \label{CPH}
\end{eqnarray}
\end{widetext}

The diagrammatic representation of the flow equation is depicted in Fig.~\ref%
{Feynman}. The dot over the four-point vertex denotes the derivative with
respect to the cutoff $\Lambda $ and the slashed loop line represents the
single-scale propagator, $S^{\Lambda}=dG_{o_{1}o_{2}}^{(0),\Lambda}/d\Lambda$.
Only one type of one-loop diagram is generated for
the P channel (PP) and C channel (PHcr). There are three one-loop diagrams for
the D channel, one corresponding to the RPA-type diagram (RPA) and the other two to the
vertex corrections (VC1, VC2). The cRPA can be reproduced by keeping only the
RPA-type series diagram (the red diagram) in the D channel. It is readily
seen that the cfRG goes beyond cRPA by including more one-loop diagrams and
these can produce significant corrections to the cRPA effective interaction.\cite{Honerkamp-2018}

In general, the two-particle interaction $V^{\Lambda }$ at the energy scale $%
\Lambda $ generated by integrating out the high energy bands during the
renormalization group flow is a function of spin indices, orbital indices,
wave vectors and Matsubara frequencies. The rich information incorporated in
this effective interaction might lead to qualitatively different physical
effects. For example, the frequency dependent interaction downfolded from
the frequency independent bare interactions was shown to be crucial to open
up a gap in undoped La$_{2}$CuO$_{4}$.\cite{Werner-2015}
On the other hand, this complexity makes it challenging to perform numerical
simulations for realistic multiband systems. To simplify the calculations, we
use the intraorbital on-site and instantaneous bilinear interaction
approximation of Ref.~\onlinecite{Honerkamp-20181} where the interaction term is
represented by
\begin{widetext}
\begin{eqnarray}
V_{o_{1}o_{2};o_{3}o_{4}}^{\Lambda }\left( k_{1},k_{2},k_{3}\right)
&=&P_{o_{1}o_{2};o_{3}o_{4}}^{\Lambda }\left( k_{1},k_{3};s\right)
+D_{o_{1}o_{2};o_{3}o_{4}}^{\Lambda }\left( k_{1},k_{4};t\right)
+C_{o_{1}o_{2};o_{3}o_{4}}^{\Lambda }\left( k_{1},k_{3};u\right)  \notag \\
&\simeq &P_{o_{1}o_{3}}^{\Lambda }\left( s\right) \delta _{o_{1}o_{2}}\delta
_{o_{3}o_{4}}+D_{o_{1}o_{2}}^{\Lambda }\left( t\right) \delta
_{o_{1}o_{3}}\delta _{o_{2}o_{4}}+C_{o_{1}o_{2}}^{\Lambda }\left( u\right)
\delta _{o_{1}o_{4}}\delta _{o_{2}o_{3}}. \label{VPDC}
\end{eqnarray}
\end{widetext}In this approximation, each channel depends on one collective
wave vector and one bosonic Matsubara frequency, $s$, $t$ or $u$, respectively.
This allows one to understand the $D$-channel as a potentially retarded interaction between density fermion bilinears. In the intraorbital onsite approximation, the two fermion operators of this bilinear have the same site, orbital and spin index, with the latter being summed over. Yet, the two bilinears coupled by this interaction in the $D$-channel can differ, i.e. $o_1 \not= o_3$ in the second line above, and hence non-local interactions are captured as well. 
The $C$- and $P$-channel terms can, in the same approximation, be interpreted as potentially retarded spin exchange and pair hopping terms, such that the effective interaction contains onsite and non-local, orbital-dependent density-density, exchange and pair-hopping interactions.  

The above equations are written in the orbital basis as the interaction part
of the three-band model is given in the orbital basis. However, it is
physically much more intuitive to describe the downfolding procedure in the
band basis. More specifically, the loop term in the particle-particle channel
can be written as

\begin{eqnarray}
&&G_{\tilde{o}_{1}\tilde{o}_{3}}^{\Lambda }\left( k\right) G_{\tilde{o}_{2}%
\tilde{o}_{4}}^{\Lambda }\left( s-k\right) =\sum_{n_{1},n_{2}}U_{\tilde{o}%
_{3},n_{1}}\left( \mathbf{k}\right) U_{\tilde{o}_{4},n_{2}}\left( \mathbf{s}-%
\mathbf{k}\right)  \notag \\
&&\hspace{10mm}\times\left[ G_{n_{1}}^{\Lambda }\left( k\right) G_{n_{2}}^{\Lambda }\left(
s-k\right) \right] U_{n_{1},\tilde{o}_{1}}^{\dag }\left( \mathbf{k}\right)
U_{n_{2},\tilde{o}_{2}}^{\dag }\left( \mathbf{s}-\mathbf{k}\right) ,\nonumber\\
\end{eqnarray}%
where $n_{1}$ and $n_{2}$ are band indices and $U$ are the unitary matrices
which diagonalize the non-interacting Hamiltonian $h_{0}\left( \mathbf{k}%
\right) $ in Eq.~(\ref{Hamiltonian}). Here, we use the flat-cutoff scheme
where the regulator simply switches off the high-energy bands irrespective
of their energy and momentum and the propagator $G_{n}^{\Lambda }\left(
k\right) $ is given by

\begin{equation}
G_{n}^{\Lambda }\left( k\right) =\left\{
\begin{array}{c}
\Lambda G_{n}^{(0)}\left( k\right) \\
G_{n}^{(0)}\left( k\right)%
\end{array}%
\right.
\begin{array}{c}
\text{for }n\in \text{high-energy bands,} \\
\text{for }n\in \text{target bands.}%
\end{array}%
\end{equation}%
The above form of the propagator excludes the contribution from the loop
term composed of two target-band propagators as the $\Lambda $ derivative
equals $0$ in this case. The effective interaction $%
V_{o_{1}o_{2};o_{3}o_{4}}^{\Lambda }$ in the orbital basis can be obtained
by integrating the flow equations from $\Lambda =1$ down to $\Lambda =0$.
Then, the effective interaction $V_{tt,tt}^{\Lambda }\left(
k_{1},k_{2},k_{3}\right) $ in the target band denoted by the band index $t$ can
be obtained by projections with the orbital-to-band transformation unitary
matrices $U$

\begin{gather}
V_{tt;tt}^{\Lambda
}(k_{1},k_{2},k_{3})=\sum_{o_{1}o_{2}o_{3}o_{4}}U_{o_{1},t}\left( \mathbf{k}%
_{1}\right) U_{o_{2},t}\left( \mathbf{k}_{2}\right)  \notag \\
\hspace{10mm}\times V_{o_{1}o_{2};o_{3}o_{4}}^{\Lambda }(k_{1},k_{2},k_{3})U_{t,o_{3}}^{\dag
}\left( \mathbf{k}_{3}\right) U_{t,o_{4}}^{\dag }\left( \mathbf{k}%
_{4}\right) .  \label{proj}
\end{gather}
The zero-frequency momentum average of this quantity with all high-energy bands removed at $\Lambda =0$ can be used to define the effective target-band onsite repulsion (see also Refs.~\onlinecite{Honerkamp-20181,Honerkamp-2018})
\begin{equation}\label{Ueff}
U_\text{eff} = \frac{1}{N^3}\sum_{\vec k_1,\vec k_2,\vec k_3} \left. V_{tt;tt}^{\Lambda = 0
}(k_{1},k_{2},k_{3})\right|_{s=u=t=0} \, .
\end{equation}
In the following, we
calculate and compare three different types of effective interactions $V_{o_{1}o_{2};o_{3}o_{4}}^{\Lambda }(k_{1},k_{2},k_{3})$ in Eq.~\eqref{proj}, which are then used in Eq.~\eqref{Ueff}: the
bare interaction, the cRPA-screened interaction and the cfRG interaction. The bare
interaction is the unrenormalized interaction projected onto the target band
using Eq.~(\ref{proj}) with band index $t$ given by the target band. The
bare interaction is Matsubara frequency independent as the orbital-to-band
transformations are only momentum dependent. Both the cRPA and the cfRG
interactions are functions of momenta and Matsubara frequencies as visible from Eq.~\eqref{VPDC}.
For the static effective onsite repulsion $U_\text{eff}$ in Eq.~\eqref{Ueff}, we set all these collective frequencies $s$, $t$ and $u$ to zero. In principle, even for onsite bare interactions, the projected effective interactions will have nonlocal terms. For the given situation these nonlocal terms will however be rather small, as can be inferred, e.g., from Ref.~\onlinecite{Honerkamp-20181}.

Besides the static $U_\text{eff}$, we discuss a frequency-dependent $U_\text{eff} (\omega_n)$ as
\begin{equation}\label{Ueffomega}
	U_\text{eff} (\omega_n) = \frac{1}{N^3}\sum_{\vec k_1,\vec k_2,\vec k_3} \left. V_{tt;tt}^{\Lambda = 0
	}(k_{1},k_{2},k_{3})\right|_{s=u=t=\omega_n} \, .
\end{equation}
Fixing the total incoming frequency $s$ and the two transfer frequencies $t$ and $u$ to the same value is of course an approximation to the full three-frequency dependence of the effective interaction. At least, this compromise, introduced in Ref.~\onlinecite{Honerkamp-20181}, captures the true frequency dependence in the limit of very small or very large values of these frequencies. Furthermore, for the QMC solution of the effective model within DMFT, this reduction is currently necessary.
Below we use DMFT(QMC) to solve the half-filled low-energy effective one-band models with the  frequency-dependent cfRG and cRPA effective interactions. The
calculations are performed on a $16\times 16$ lattice at a temperature $%
T=0.1$. 

\section{Results}

\label{results} In this section, we calculate the effective interactions in
the low-energy target band for the three-band Emery model with cRPA and cfRG
as a function of the charge-transfer energy $\Delta _{dp}$ and the interatomic $2p$-%
$3d$ interaction $U_{pd}$, in both undoped or doped situations. We begin
the discussion with the results for the undoped state, which corresponds to
each CuO$_{2}$ unit cell occupied by five electrons. We denote this situation
by $n=1$ in the following for simplicity. Subsequently, the electron-doped ($n>1$)
and hole-doped ($n<1$) cases will be discussed.

\subsection{Frequency and momentum dependence of the effective interactions}
\label{sec_frequency}

\begin{figure}[tbp]
\includegraphics[width=0.99\columnwidth]{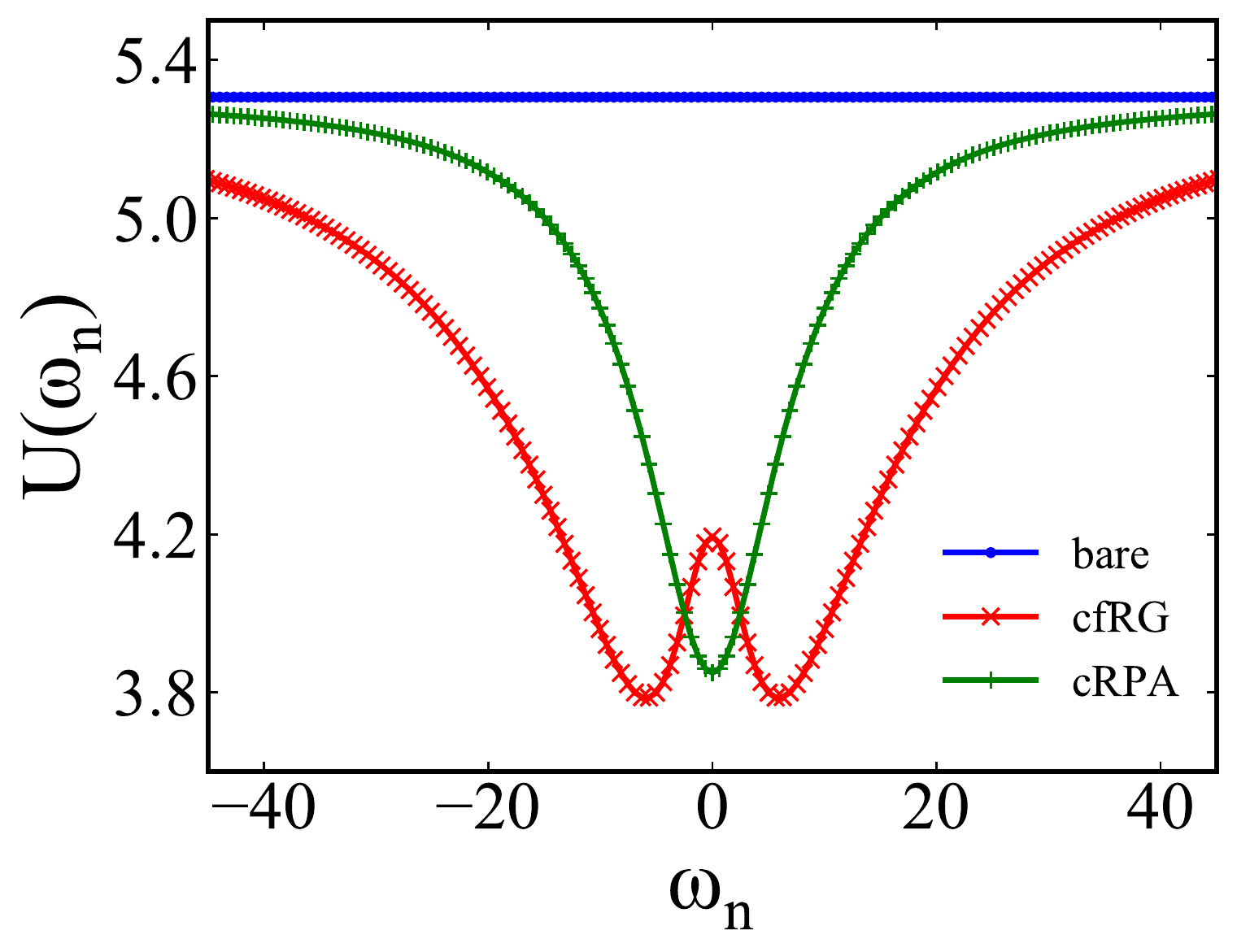} %
\includegraphics[width=0.99\columnwidth]{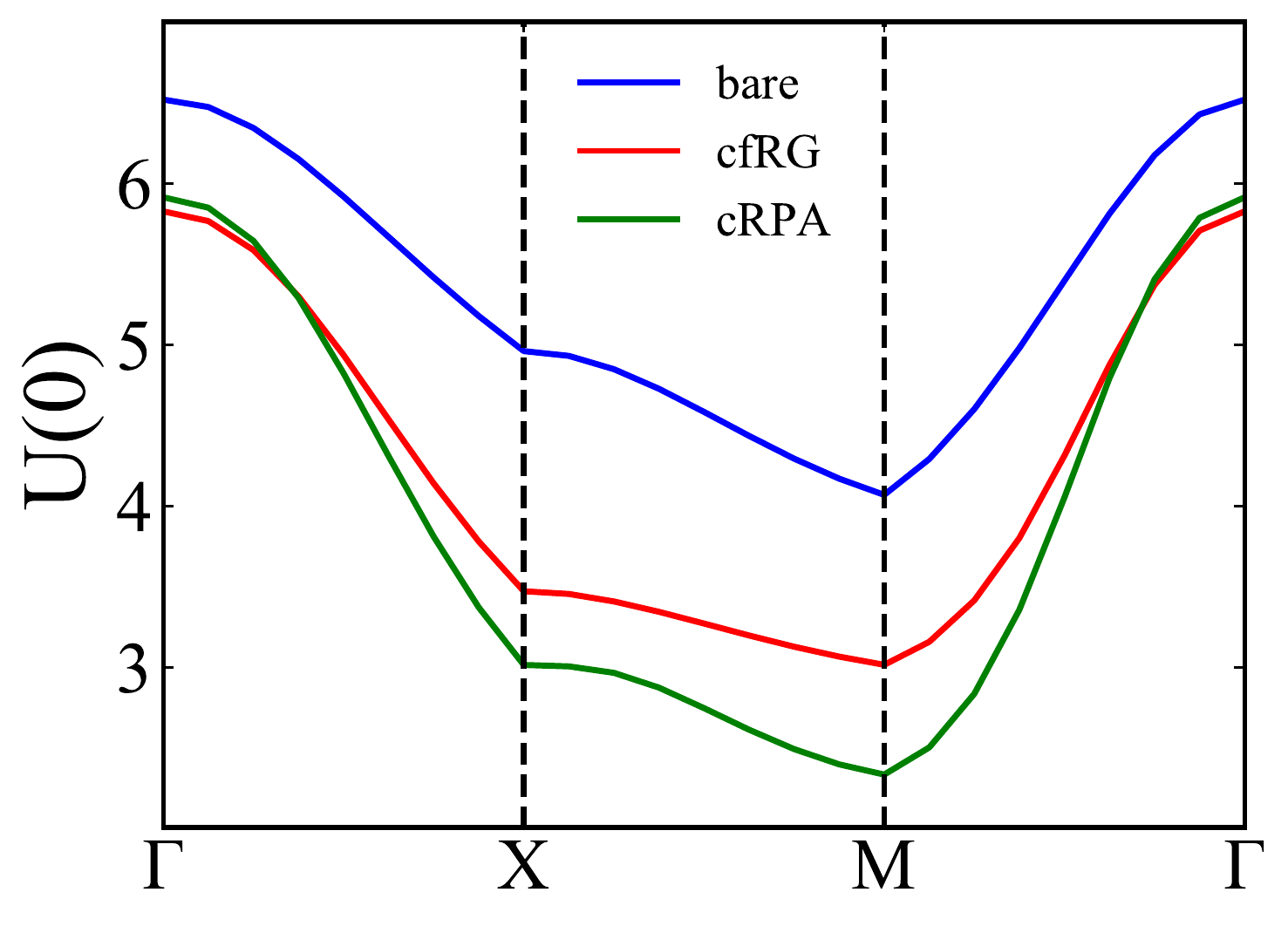}
\caption{Upper panel: The frequency dependence of the effective onsite (i.e. momentum averaged) interaction in the target band at the Fermi level.
Lower panel: The momentum dependence of the static (zero frequency) effective interaction in the target band. All values are given in units of eVs.}
\label{Uef}
\end{figure}

As can be seen from the cfRG flow equations, Eq.~(\ref{cfrg_flow}), the
downfolded effective interaction will develop rich frequency and momentum
structures even for static bare interactions. We show the momentum
averaged effective interactions $U_\text{eff}$ as defined in Eq.~(\ref{Ueff}) as a function of Matsubara
frequency for $U_{pd}=2$, $\Delta _{dp}=2$ at half filling in the upper panel
of Fig.~\ref{Uef}. The bare effective interaction without loop corrections
is Matsubara frequency independent but lower than the $U_{dd}$ interaction in the three-band model, due to the transformation (\ref{Ueff}). The cfRG and cRPA results are nontrivial functions of Matsubara frequency $\omega_n$.
In the static limit, cfRG and cRPA screen down the bare interaction to less than 80\% and this screening
effect is slightly stronger in cRPA than that in cfRG.
Initially, the cfRG effective interaction is decreasing with increasing $\omega_n$, while the cRPA effective interaction is increasing.  However, the frequency dependence of the cfRG effective interaction is non-monotonic and exhibits an upturn around $\omega_n\approx 13$. This non-monotonic behavior of the cfRG interaction is expected, as screening becomes ineffective at high frequencies and both the cfRG and cRPA effective interactions should approach the bare interaction. The maximum at zero frequency in the cfRG interaction can be understood as an effect of the antiscreening $C$-channel that mixes into the charge channel most strongly at small frequencies.
The frequency-dependent effective interaction has been argued to play a crucial role in opening the gap for the insulating state of La$_{2}$CuO$_{4}$\cite%
{Werner-2015} and it shares similar features within the cuprate family.\cite%
{Nilsson-2019} Note that the analytic continuation from Matsubara to real
frequencies remains a challenging task and is left for future work.

The lower panel of Fig.~\ref{Uef} illustrates the momentum dependence of the
zero frequency interactions along the high-symmetry directions of the
Brillouin zone. All of the three interactions have a strong momentum
dependence with the maximum values at the $\Gamma $ point. As $k$ moves from
$\Gamma = (0,0)$ to $X=(\pi,0)$ and $X$ to $M=(\pi,\pi)$, 
the interactions decrease and reach the
minimum value at $M$ point. Along $M$ to $\Gamma $, the opposite
behavior is observed. Compared with the bare interaction, the suppression is
stronger along $X$ to $M$ and becomes smallest at the $\Gamma $ point. The
overall suppression of the static value is weaker in cfRG than in cRPA.

One main take-away message from these figures is that the differences between the frequency- and momentum dependences of the three interaction types, bare, cRPA and cfRG, is less prominent that the overall suppression effects in cRPA and cfRG. Hence, in a first analysis one can concentrate on the local, zero-frequency effective interaction, the `effective $U$'.

\subsection{Undoped case for different $\Delta _{dp}$ with fixed $U_{dp}$}

\begin{figure}[tbp]
\includegraphics[width=0.99\columnwidth]{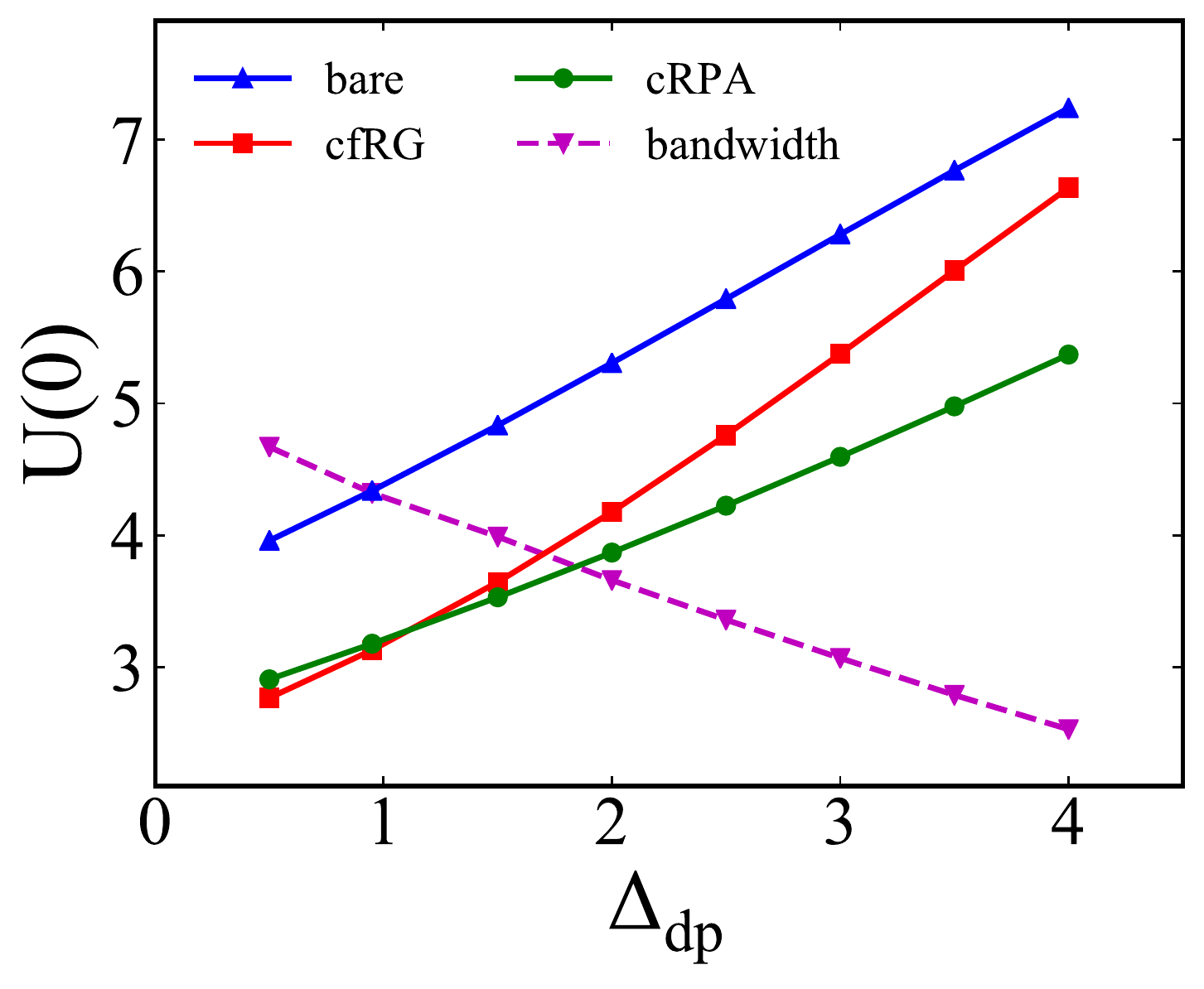} %
\includegraphics[width=0.95\columnwidth]{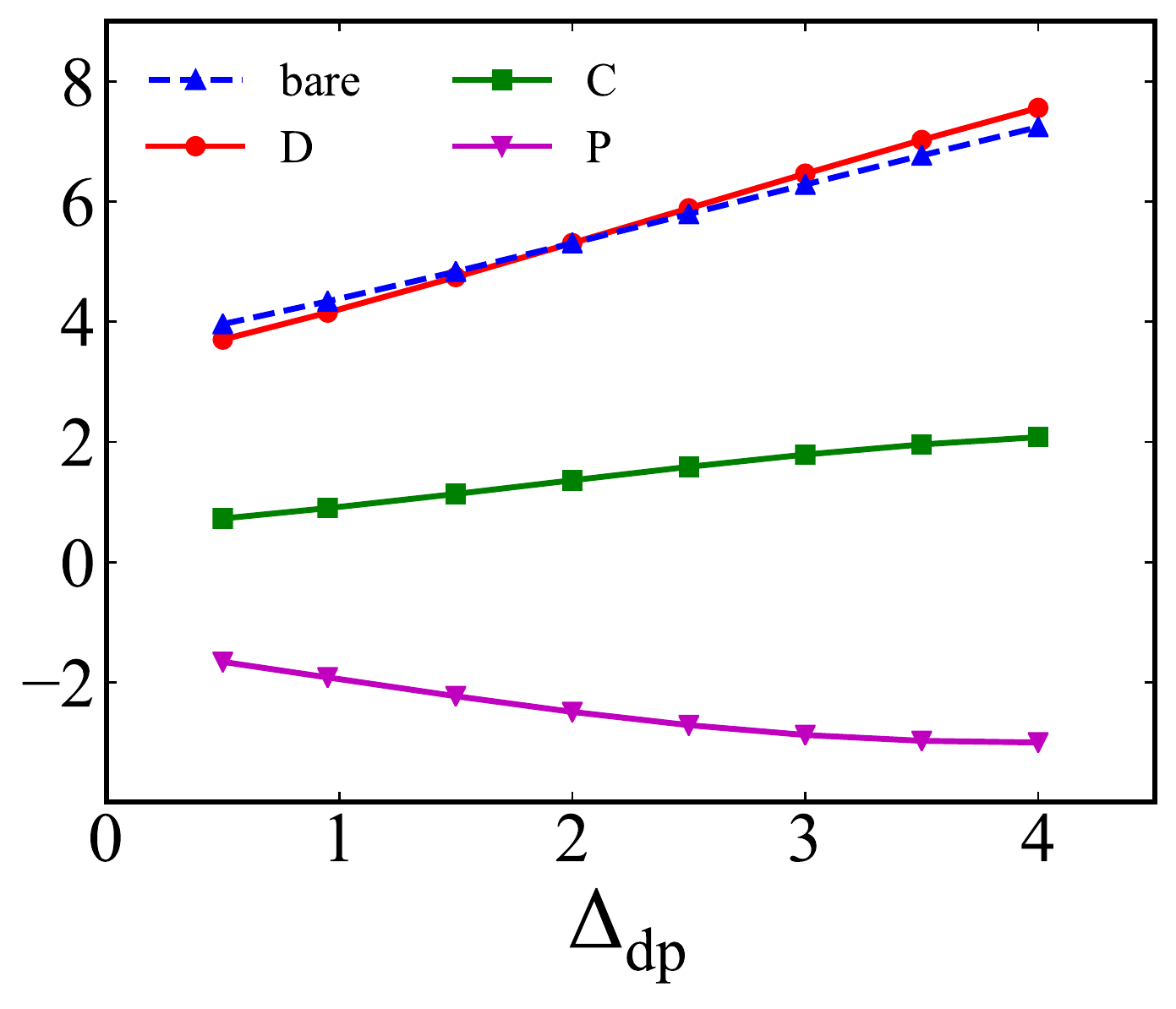}
\caption{Upper panel: The effective interaction and the bandwidth of the target band as a function of $\Delta
_{dp}$. Lower panel: The channel coupling functions as a function of $\Delta
_{dp}$. Data for $U_{dd}=13$.}
\label{ctg}
\end{figure}

The charge-transfer energy $\Delta _{dp}$ is one of the key parameters in
the Emery model and the value of $\Delta _{dp}$ varies substantially between
different cuprate families. The strength of the $d$-$p$ hybridization and
the charge (spin) distribution on the CuO$_{2}$ layer are controlled by the
value of $\Delta _{dp}$. The physical properties of the Emery model have
been found to vary fundamentally with $\Delta _{dp}$ in the underdoped regime%
\cite{Chiciak-2018} and even at half-filling.\cite{Vitali-2019} Moreover,
both theoretical and experimental results suggest that $\Delta _{dp}$ is
anticorrelated with the maximal transition temperature $T_{c,\max }$ in
cuprates,\cite{Weber-2012,Ruan-2016,Rybicki-2016,Zegrodnik-2019} that is,
reducing $\Delta _{dp}$ yields a larger $T_{c,\max }$. However, it is
difficult to determine the accurate value of $\Delta _{dp}$ from \textit{%
ab initio} calculations. As mentioned before, the gap between the low-energy
target band and the high-energy screening bands is roughly given by $\Delta
_{dp}$. Thus, it is interesting to investigate the screening effect of the
Emery model as a function of $\Delta _{dp}$. In this subsection, we set $%
U_{dp}=2$ and scan the charge-transfer energy $\Delta _{dp}$.

From the upper panel of Fig.~\ref{ctg}, one can clearly see that the
screening is slightly stronger for cfRG compared to cRPA when $\Delta _{dp}<1
$ and becomes weaker as $\Delta _{dp}$ increases. As the values of $\Delta
_{dp}$ in most of the cuprates families are larger than 1,\cite{Vitali-2019}
overscreening of the static $U$ in the cRPA downfolding scheme is expected for realistic \textit{%
ab initio} calculations. Screening is less effective for larger $\Delta
_{dp} $ in cfRG, where the effective interaction is screened down to 70\% of
the bare interaction for the smallest $\Delta _{dp}$ and 92\% of the bare
interaction for the largest $\Delta _{dp}$ used in this study. This is
consistent with the physical picture that a stronger correlation for the
target band is obtained with a larger $\Delta _{dp}$ as the $d$-$p$
hybridization is weaker. However, the ratio of the cRPA effective
interaction to the effective bare interaction ($U_\text{eff}^\text{cRPA}/U_\text{bare}$) for
different $\Delta _{dp}$ remains almost unchanged. Note however that larger $\Delta _{dp}$
also implies a narrower conduction band (see Fig.~\ref{ctg}), i.e. the effective correlation strength 
increases even more with $\Delta_{dp}$ than what is suggested by the increase of $U(0)$.

According to Eq.~(\ref{cfrg_flow}), the final cfRG interaction can be
decomposed into the sum of three channels (D, C and P). To gain further
insights into the cfRG results, we show the three channel coupling functions
and the bare interactions in the lower panel of Fig.~\ref{ctg}. From this
plot we see that both the direct and crossed particle-hole coupling
functions are positive while the particle-particle coupling functions are
negative for all the $\Delta _{dp}$ studied. With increasing $\Delta _{dp}$,
the absolute values of the three channel coupling functions increase. The
direct particle-hole coupling function is approximately equal to the bare
interaction, which implies that the RPA contribution (RPA in Fig.~\ref%
{Feynman}) to the screening is almost canceled by the vertex corrections
(VC1 and VC2 in Fig.~\ref{Feynman}). This near-cancellation behavior has also
been found in Ref.~\onlinecite{Honerkamp-2018}. Since the particle-particle coupling
function is more negative than the positive crossed particle-hole
coupling function, the bare interaction is screened down in cfRG.

\subsection{Undoped case for different $U_{dp}$ with fixed $\Delta _{dp}$}

\begin{figure}[tbp]
\includegraphics[width=0.99\columnwidth]{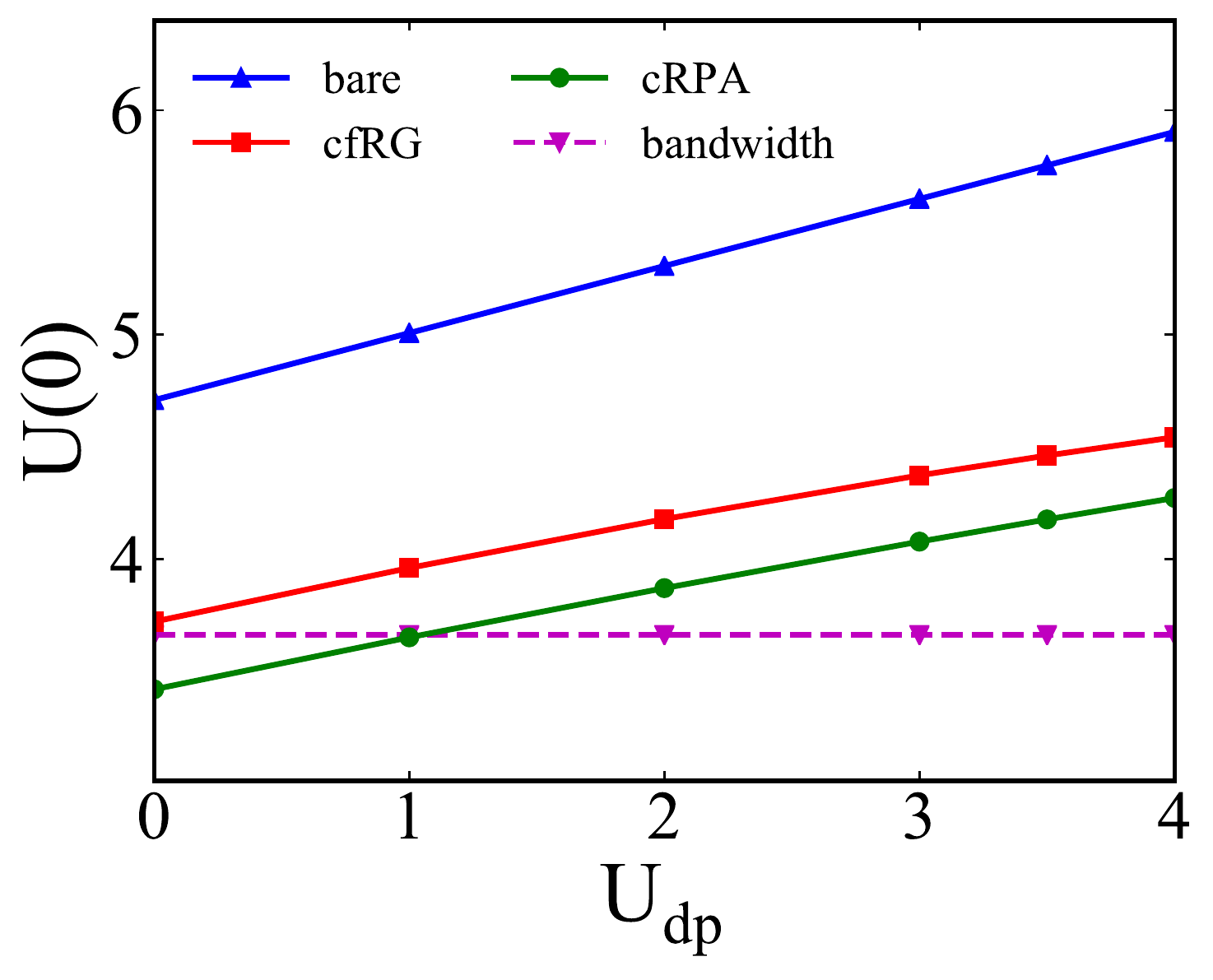} %
\includegraphics[width=0.95\columnwidth]{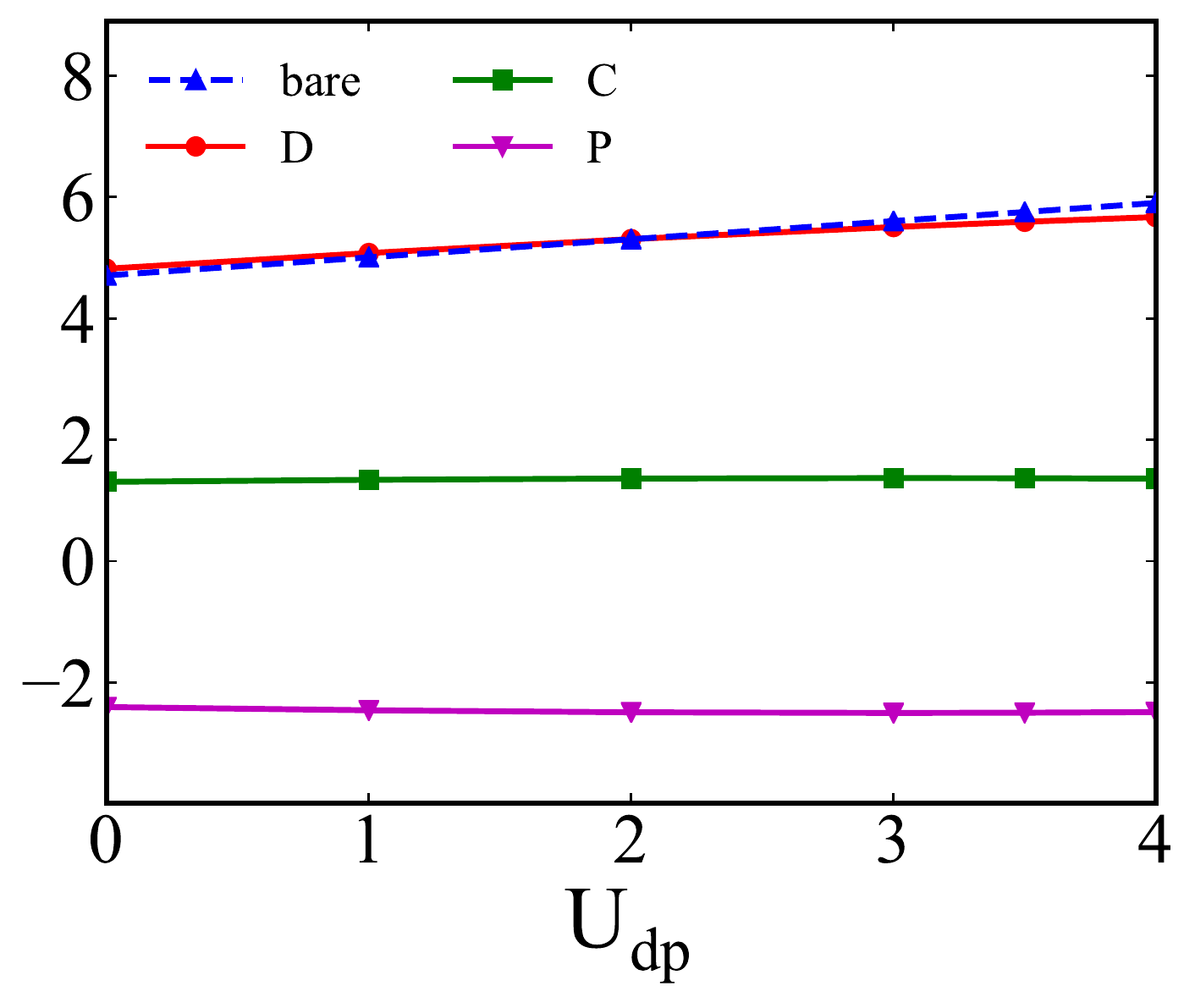}
\caption{Upper panel: The effective interaction and the bandwidth of the target band as a function of $U_{dp}$.
Lower panel: The renormalized channel coupling functions as a function of $%
U_{dp}$.}
\label{Udp}
\end{figure}

The interatomic $2p$-$3d$ interaction $U_{dp}$ in the Emery model is often
neglected because $U_{dp}$ is small compared to $U_{dd}$. However, detailed
numerical calculations have been performed to argue that in DMFT calculations, the inclusion of $%
U_{dp}$, at least at the Hartree level, is crucial to
drive a metal-insulator transition of the charge-transfer type for the
Emery model.\cite{Hansmann-2014,Werner-2015} Furthermore, excitonic
fluctuations in the doped cuprates\cite%
{Hirayama-2019} and photo-induced bandshifts\cite{Golez-2019} may also be affected by $U_{dp}$.
In this subsection, we discuss the momentum averaged static
effective interactions $U_\text{eff}\left( \omega _{n}=0\right) $ as a function
of $U_{dp}$ at half filling for fixed charge-transfer energy $\Delta
_{dp}=2 $.

The results are presented in Fig.~\ref{Udp}. Here, the reduction of the bare
interaction is stronger for cRPA. In contrast to the $\Delta
_{dp}$ dependence, in both cases, the ratios of the downfolded effective interaction to
the bare effective interaction for different $U_{dp}$ are almost the same.
Also the width of the conduction band does not depend on $U_{dp}$. 

The overall changes of the effective interactions as a function of $U_{dp}$
are less significant compared to the variation with $\Delta _{dp}$. Our
data indicate that larger $U_{dp}$ corresponds to stronger correlations in
the low-energy target band.
Together with the DMFT result that a charge gap
opens in the one-band Hubbard model if the effective interaction goes beyond some critical value,\cite{Georges-1996}
 the trend found in Ref.~\onlinecite%
{Hansmann-2014} is consistent with our results. In the lower panel of Fig.~%
\ref{Udp}, the three channel coupling functions together with the bare
interactions are given. Again, the near-cancellation of the direct
particle-hole coupling function is observed. However, the coupling functions
in the crossed particle-hole and particle-particle channels are nearly
independent of $U_{dp}$.

\subsection{Doped case for fixed $U_{dp}$ and $\Delta _{dp}$}

\begin{figure}[tbp]
\includegraphics[width=0.99\columnwidth]{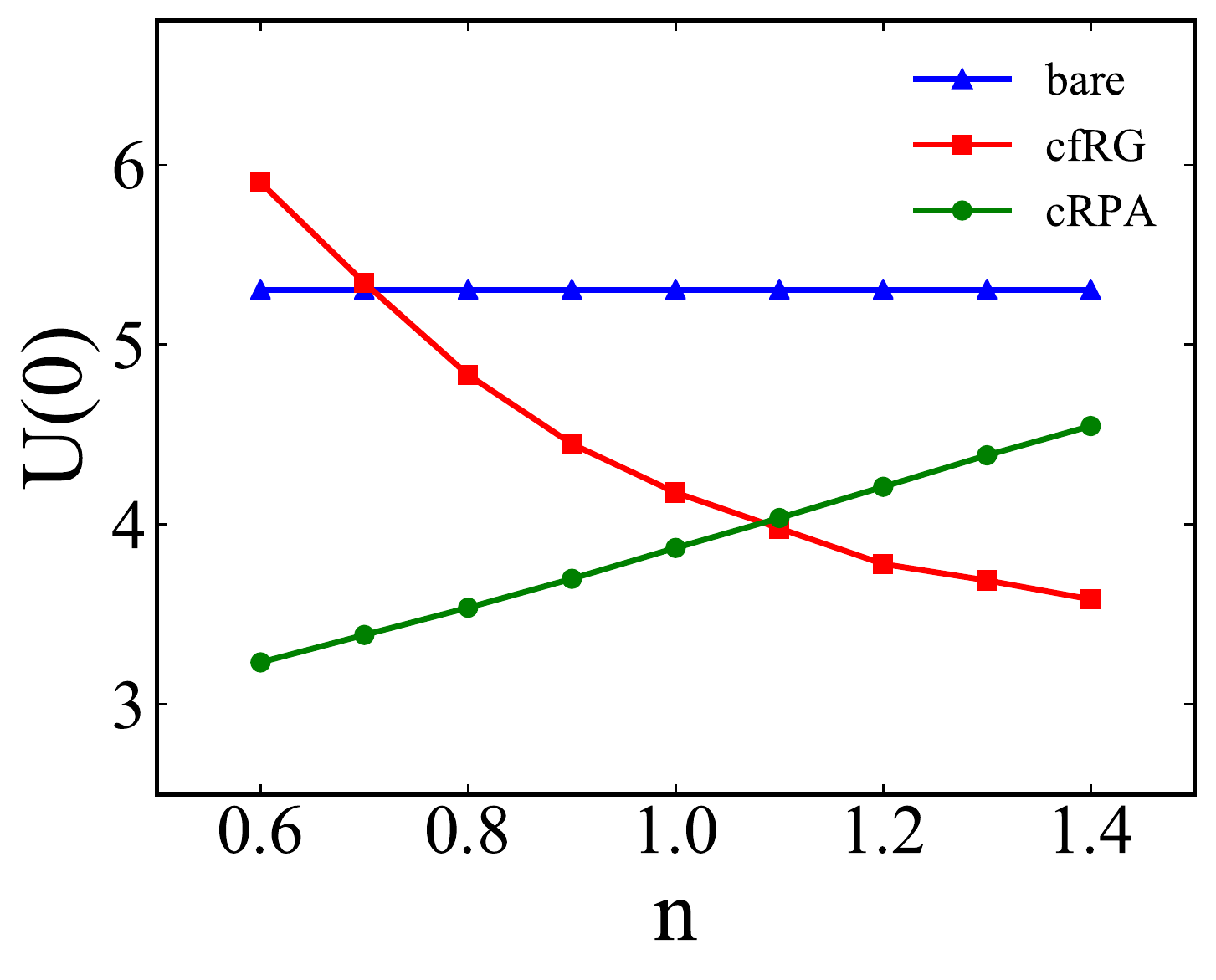} %
\includegraphics[width=0.99\columnwidth]{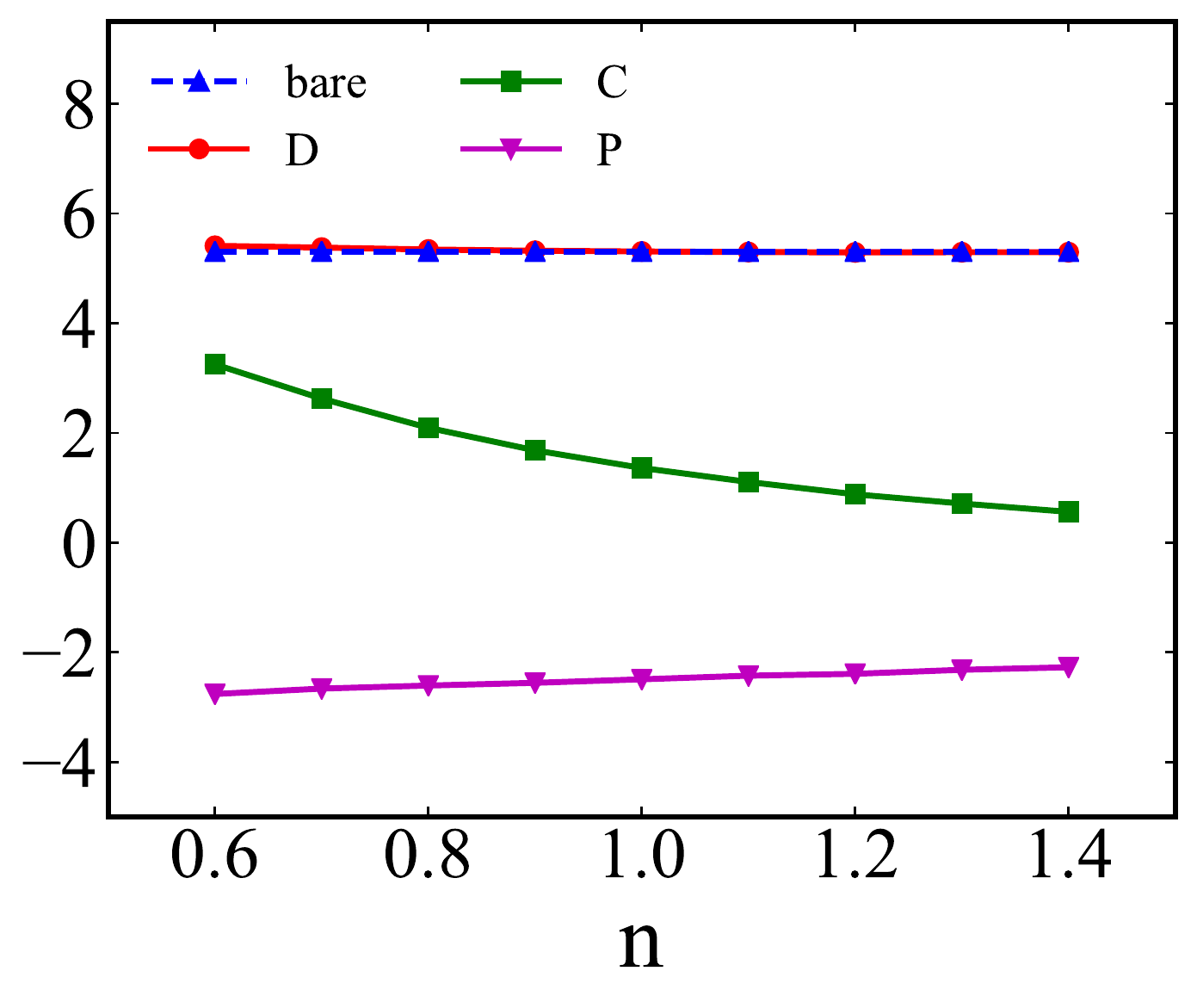}
\caption{Upper panel: The effective interaction as a function of doping.
Lower panel: The renormalized channel coupling functions as a function of
doping. Here we set set $U_{dp}=2$ and $\Delta_{dp}=2$. 
}
\label{doping}
\end{figure}

The doped system is more interesting, since many of the correlation driven exotic
phases (pseudogap, strange metal and various density wave states) have been
proposed and observed in the doped cuprates.\cite%
{Lee-2006,Keimer-2015,Fradkin-2015} In many numerical studies, the model
parameters are treated as independent of doping, however, it should be noted
that the values of $\Delta _{dp}$ and $U_{dp}$ may change upon doping due to
the shift of the O 2$p$ and Cu 3$d$ states.\cite{Xiang-2009} Accordingly,
the values of the effective interactions should change, as discussed in
the previous subsections for the undoped cases. To simplify the discussion and
focus on the changes of the effective interactions originating from doping, we
ignore this effect in this subsection and set $U_{dp}=2$ and $\Delta _{dp}=2$.

We show in Fig.~\ref{doping} the doping dependence of the momentum averaged
static effective interactions and the channel decomposed results. Obviously,
the bare interaction remains constant as a function of doping since the
orbital-to-band transformation matrices are independent of band-filling. In
cfRG, the correlation strength is suppressed with increasing filling $n$,
and the crossed particle-hole channel is responsible for this
decrease of the effective interaction. This suggests that the correlations are
stronger in the hole-doped system compared with the electron-doped system. Similar
to our results for the undoped cases, the near-cancellation in the
direct particle-hole channel is also found here. Contrary to the crossed
particle-particle channel, the effective interaction in the
particle-particle channel slightly increases with doping. Besides the direct
particle-hole channel, the contributions from the crossed particle-hole channel
outweigh the particle-particle channel for $n<0.7$, which results in the
antiscreening effect. Here, the cRPA results follow an opposite trend compared to
cfRG as a function of doping. The screening effects become weaker with
increasing $n$. The two downfolded interaction curves cross near $%
n=1.1$, below which the effective interaction is stronger for cfRG. In Refs.~\onlinecite{Nilsson-2019,Jang-2016},
the cRPA results suggest that most
electron-doped cuprates have a smaller $U$ compared to the hole-doped
ones. At first sight, this seems to contradict with our results. However,
their results are obtained by doping different parent compounds, which means
that different model parameters are used during the calculation. It still remains
an experimental challenge to systematically scan the phase diagram from hole
doping to electron doping for a single material.\cite{Orenstein-2010}

\begin{figure*}[t]
\includegraphics[width=0.65\columnwidth]{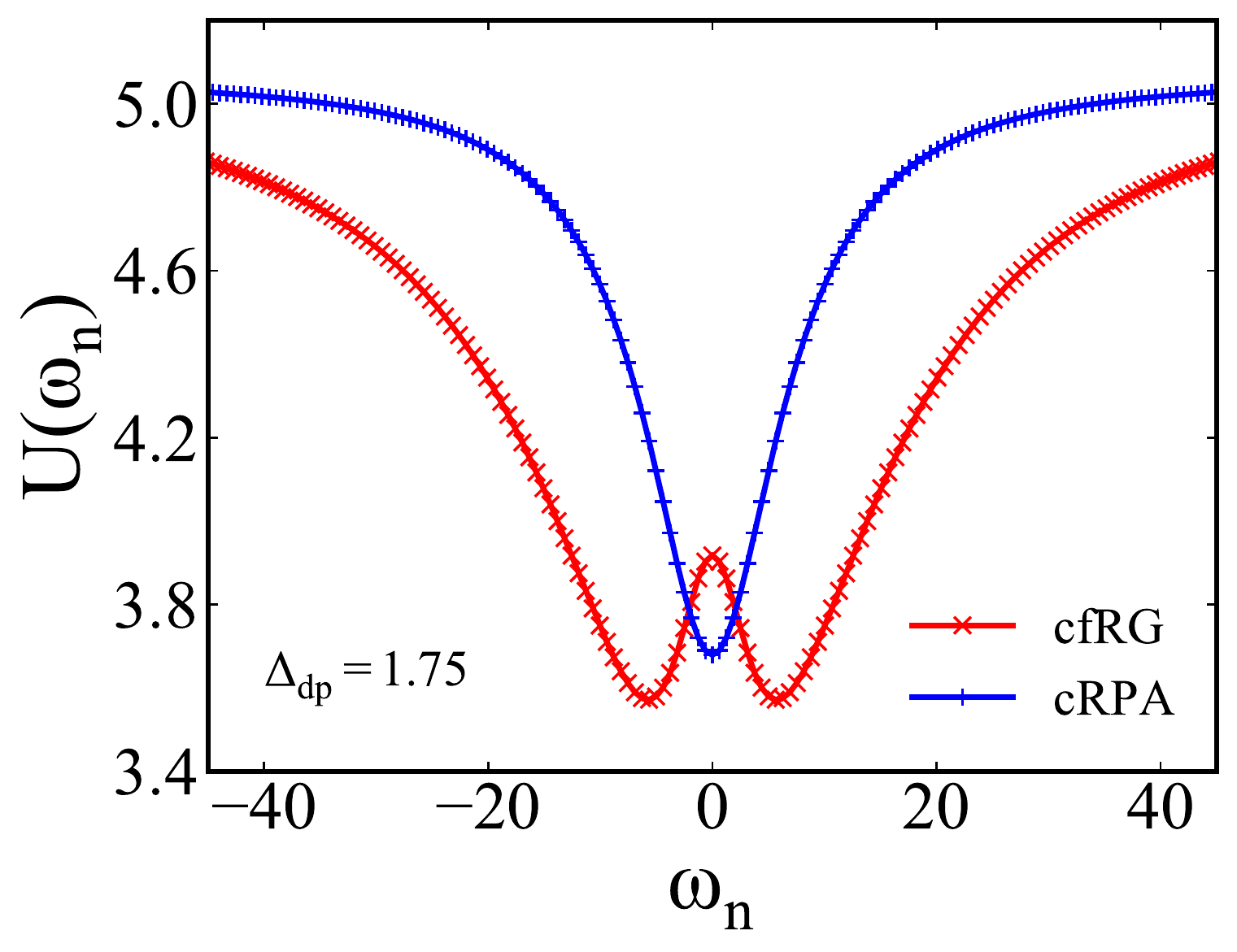} %
\includegraphics[width=0.65\columnwidth]{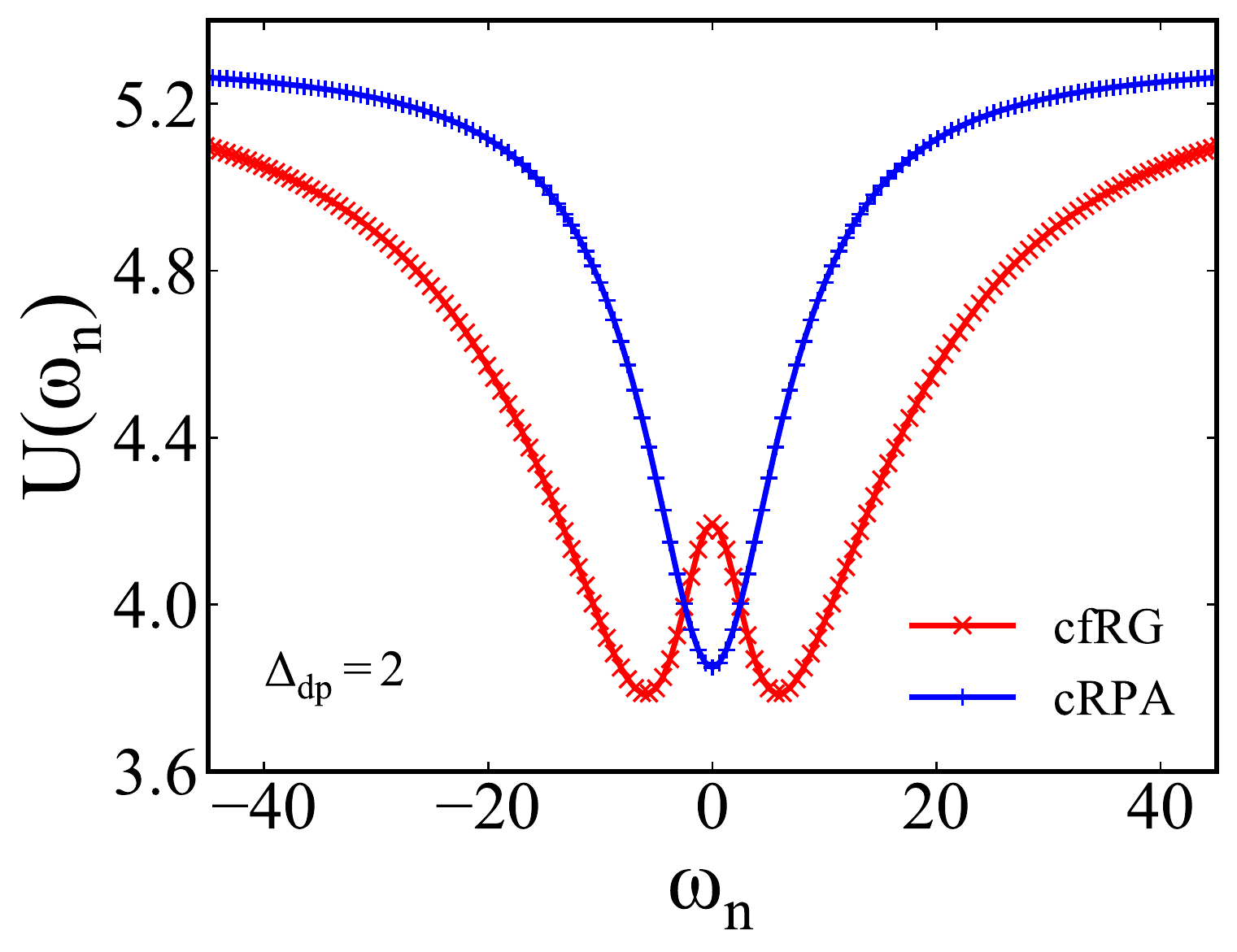}
\includegraphics[width=0.65\columnwidth]{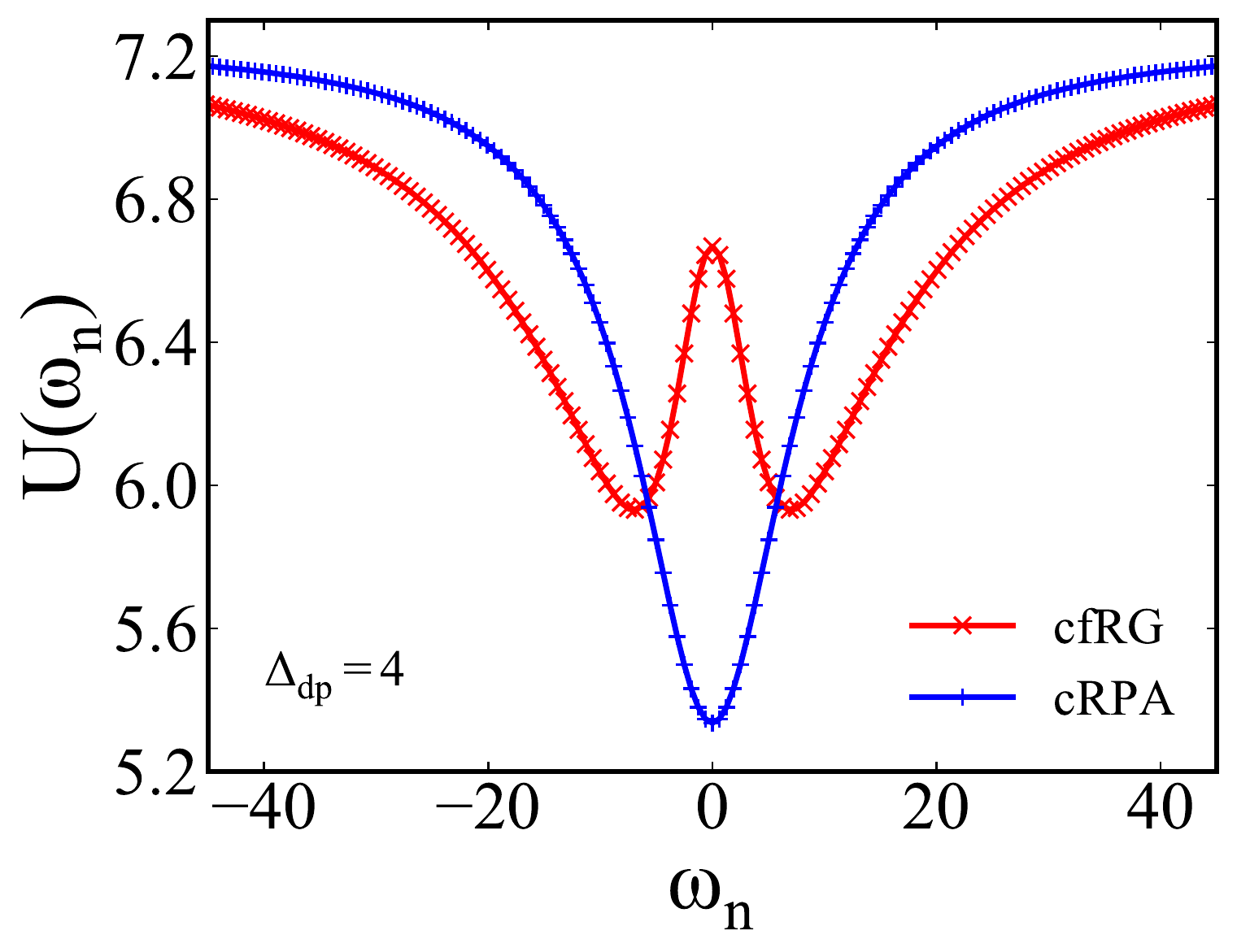}\\
\includegraphics[width=0.65\columnwidth]{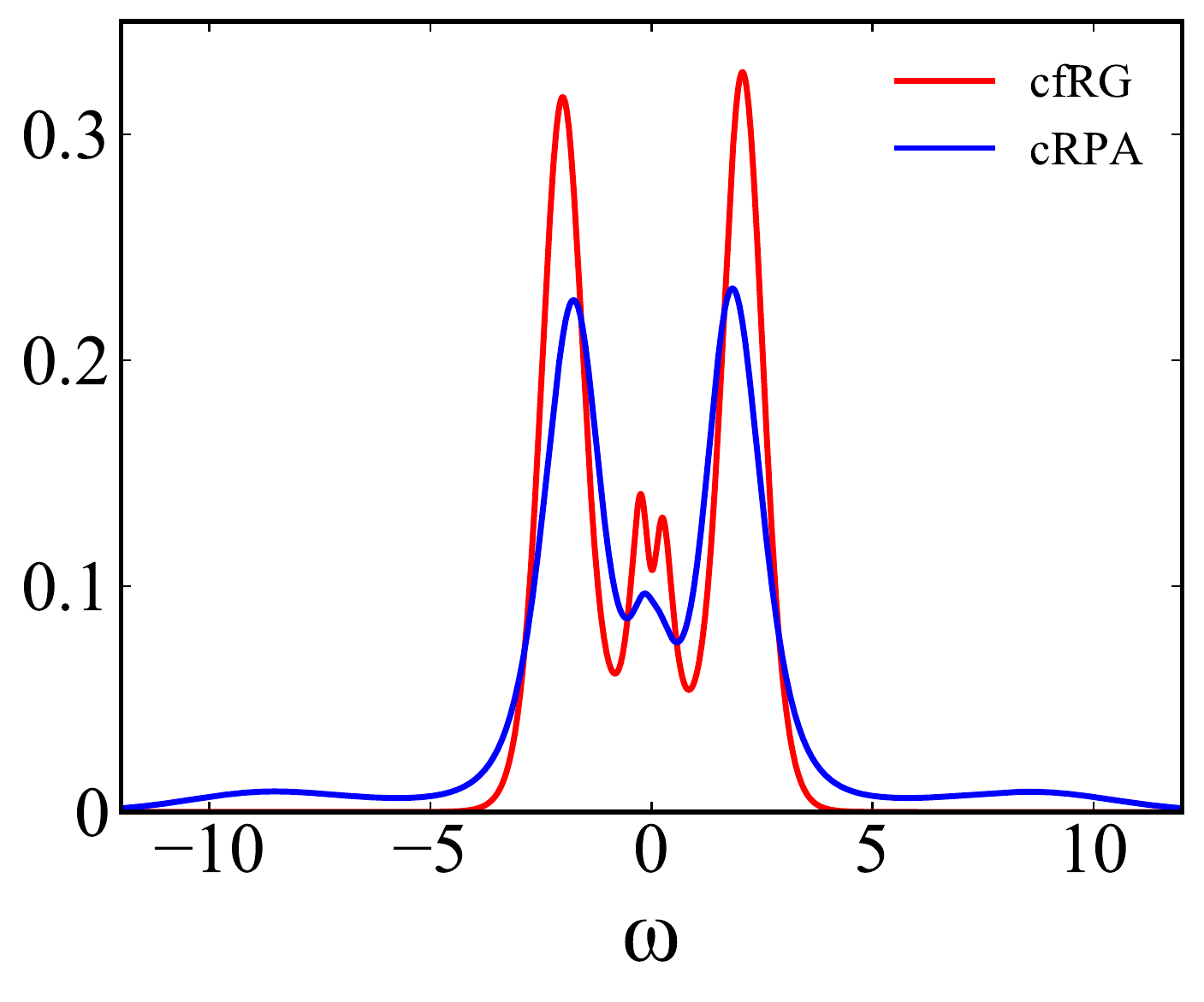} %
\includegraphics[width=0.65\columnwidth]{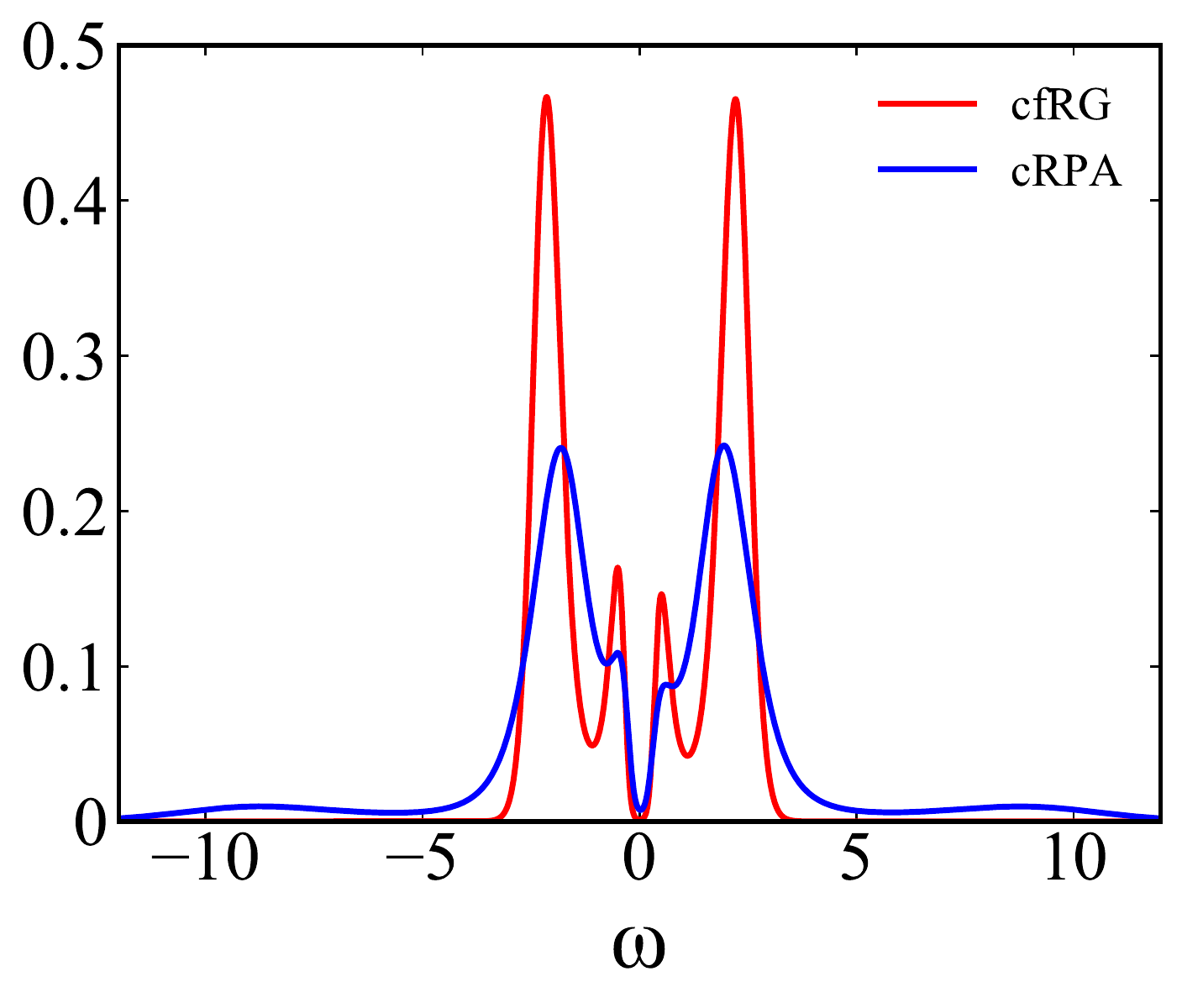}
\includegraphics[width=0.65\columnwidth]{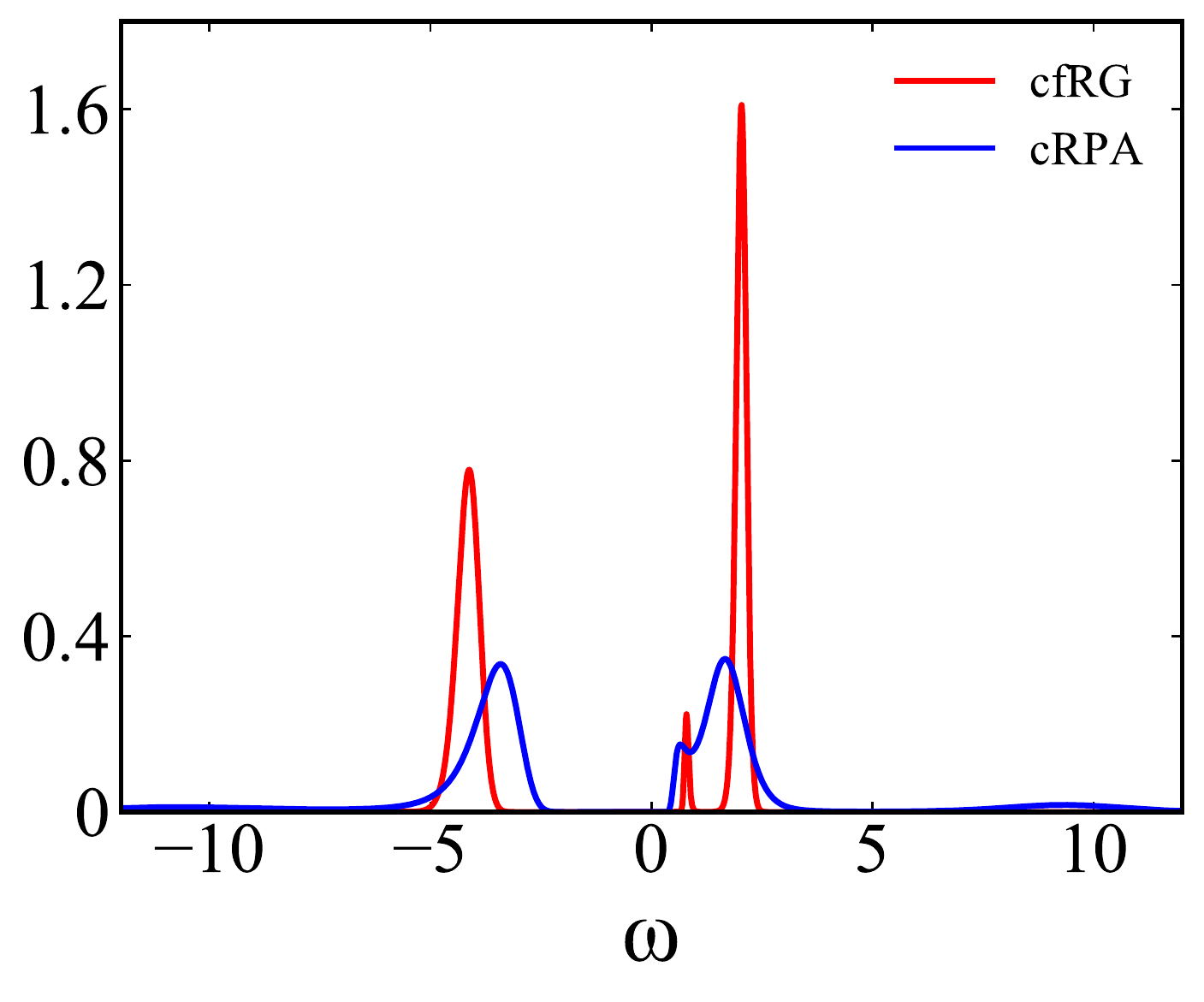}
\caption{Effective interactions (top panels) and local spectral functions (bottom panel) for the downfolded single-band models with $\Delta_{dp}=1.75$, $2$, and $4$ (from left to right).
}
\label{fig_dmft}
\end{figure*}

\section{DMFT solution of the low-energy model}
\label{DMFT}

If the interactions are restricted to the local density-density component, the downfolded single-band
models with frequenency dependent $U_\text{eff}(\omega_n)$
can be solved with DMFT using the techniques developed in Refs.~\onlinecite{Werner-2007,Werner-2010}.
In DMFT, the lattice
 system is mapped to an impurity model with a self-consistently determined bath of noninteracting
 electrons (or hybridization function).\cite{Georges-1996} An efficient method for solving this impurity model is the hybridization expansion continuous-time Monte Carlo technique,\cite{Werner-2006} which expands the partition 
 function of the impurity model in powers
 of the hybridization function and stochastically samples the corresponding diagrams. In the case of frequency-dependent interactions, all the fermionic creation and annihilation operators in these diagrams are linked by a bosonic function
 $K(\tau)$, which is the twice integrated retarded interaction,\cite{Ayral-2013,Werner-2016}
 \begin{equation}
 K(\tau) = \frac{1}{\beta}\sum_{n\ne 0}\frac{U_\text{eff}(\omega_n)-U_\text{eff}(0)}{(i\omega_n)^2}(e^{-i\omega_n\tau}-1),
 \end{equation}
while the instantaneous interaction is given by the static value $U_\text{eff}(0)$. The output of the impurity solver is the impurity Green's function $G$, which at self-consistency becomes the DMFT approximation to the local lattice Green's function.

\begin{table}
\begin{tabular}{c | c | c | c }
$\Delta_{dp}$ & $U_\text{eff}(0)$ cfRG  & $U_\text{eff}(0)$ cRPA & bandwidth \\
\hline
1.75	& 3.92 & 3.68 & 3.83\\
2 & 4.19 &  3.85 & 3.69\\
2.5 & 4.78 & 4.20 & 3.35 \\
3 & 5.41 & 4.57 & 3.08 \\
4 & 6.67 & 5.34 & 2.54
\end{tabular}
\caption{Static interactions and bandwidths of the downfolded models for different $\Delta_{dp}$.}
\label{tab_u}
\end{table}

In Fig.~\ref{fig_dmft} we plot $U(\omega_n)$ and the resulting DMFT spectral functions $A(\omega)=-\frac{\text{Im}G(\omega)}{\pi}$, obtained with Maximum Entropy analytical continuation,\cite{lewin_maxent} for three different values of $\Delta_{dp}$. For $\Delta_{dp}=1.75$, the model is metallic, while for $\Delta_{dp}=2$ it is a small-gap insulator and for $\Delta_{dp}=4$ a large-gap insulator. This metal-insulator transition is a consequence of the $\Delta_{dp}$-dependence of $U_\text{eff}(\omega_n)$ and the bandwidth of the low-energy model (Tab.~\ref{tab_u}). It becomes clear though from the upper
panels of Fig.~\ref{fig_dmft} that the static values of $U_\text{eff}$ are not sufficient to quantify the interaction effects, because of the frequency dependence, which is qualitatively different for cfRG and cRPA downfolding. For example, the insulator-metal transition happens at almost the same critical $\Delta_{dp}$ for the cfRG and cRPA interactions, despite their substantially different static values, because of the opposite low-energy screening properties (see also Sec.~\ref{sec_frequency}). While the static cfRG interaction is larger than the static cRPA interaction, it initially decreases as a function of $\omega_n$, as a result of low-energy anti-screening processes, which are not captured by cRPA. The latter downfolding procedure leads by construction to a monotonically increasing $U_\text{eff}(\omega_n)$. Hence, the effective interaction strength in the two downfolded models is comparable, which leads to similar critical $\Delta_{dp}$, while the shape of the spectral functions is different.

\begin{figure}[tbp]
\includegraphics[width=0.99\columnwidth]{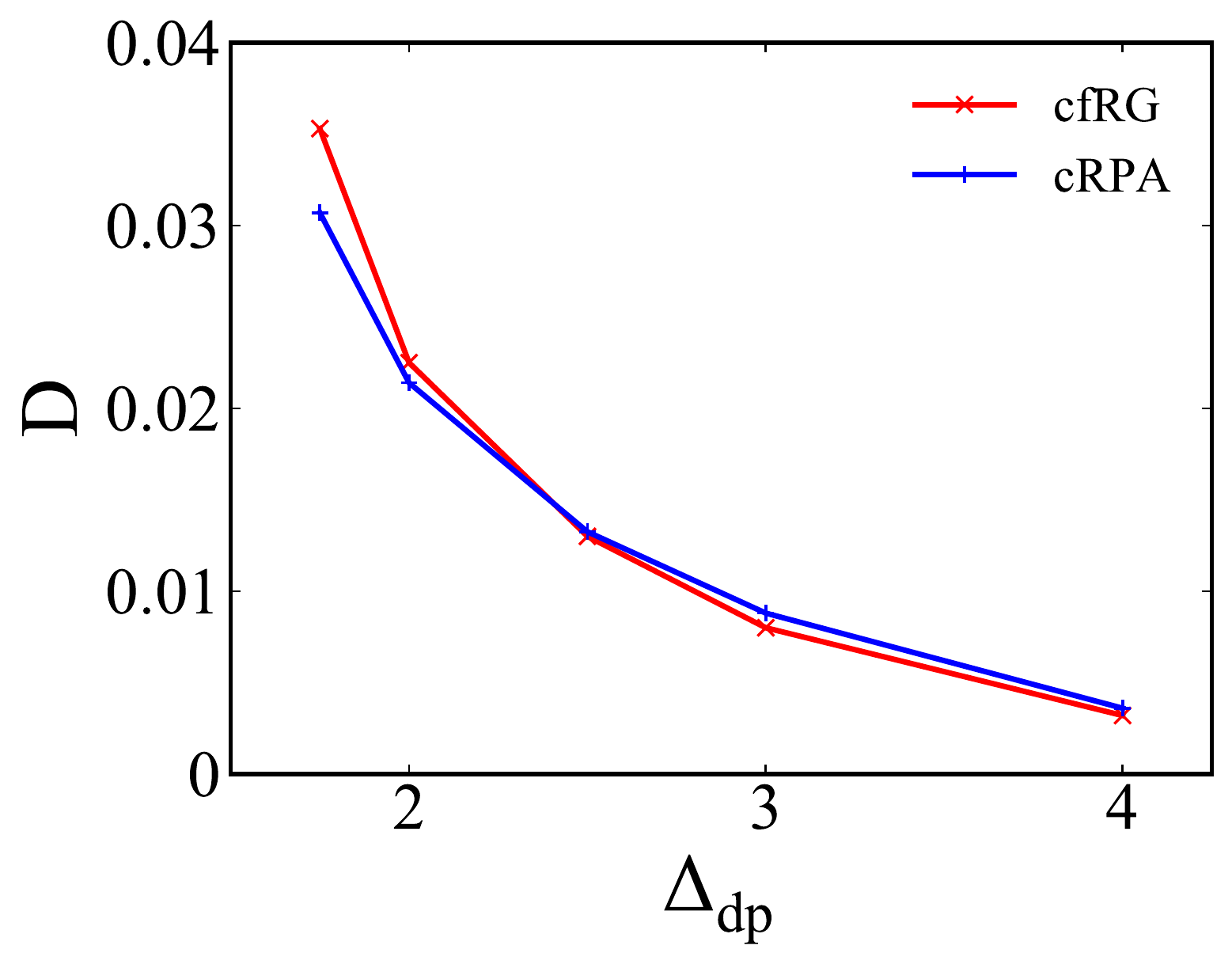}
\caption{Double occupancy $D$ as a function of $\Delta_{dp}$.}
\label{Double}
\end{figure}

A direct measure of the effective interaction strength is the average double occupation $D=\langle n_\uparrow n_\downarrow\rangle$, which is plotted as a function of $\Delta_{dp}$ in Fig.~\ref{Double}. These results confirm that the cRPA and cfRG interactions, if the full frequency dependence is taken into account, produce similar correlation effects. According to this measure, the cfRG interaction is effectively weaker than the cRPA interaction in the vicinity of the insulator-metal transition, despite the larger static values, while it is effectively stronger for $\Delta_{dp} \gtrsim 2.5$.

The frequency dependence of the interaction contains information on the screening processes involving the $p$ bands, which propagates into the DMFT solution of the low-energy model.
In particular, the spectra obtained with the cRPA interaction feature broad satellites at an energy comparable to the $d$-$p$ splitting in the original bandstructure, as has been discussed in the context of previous DMFT+$U_\text{cRPA}(\omega)$ studies.\cite{Werner-2012,Werner-2015} The spectrum for the cfRG interaction does not exhibit such satellites, presumably because of the qualitatively different frequency structure with an anti-screening peak around $\omega=5$ in the real-frequency spectrum $\text{Im}U_\text{eff}(\omega)$.

\section{Summary}

\label{conclusion}

We have presented a study of the downfolded effective interactions for the
Emery model by comparing cRPA and cfRG schemes. The momentum and frequency
structure was studied and we calculated the effective interactions by
scanning the values of the charge transfer gap $\Delta _{dp}$ and the interatomic $%
2p$-$3d$ interaction $U_{pd}$ at half-filling. By including all five
one-loop diagrams in cfRG, we have found significant corrections to cRPA
effective interactions for some parameter sets. The static interaction was generically found
to be overscreened by cRPA, which suggests that including
other one-loop terms and vertex corrections can lead to different
predictions. According to our data, the effective interaction
increases as $\Delta _{dp}$ increases, and the trend is similar as a function
of $U_{pd}$ if one of the parameters is fixed. This
indicates that the charge transfer insulating state is stabilized by a
larger $\Delta _{dp}$ or $U_{pd}$, which is compatible with the results
obtained in Ref.~\onlinecite{Hansmann-2014}. The effective
interaction was found to be more sensitive to $\Delta _{dp}$ than $U_{pd}$.

We also studied the doping dependence of the effective interactions. Away from
half-filling, the cfRG interaction decreases with increasing particle
number mainly due to the decrease of the crossed particle-hole channel,
while the opposite trend is observed for cRPA. An antiscreening effect
which cannot be captured by cRPA, is found by cfRG in the hole-doped case.
For all the cases studied in this paper, a near-cancellation of the direct
particle-hole channel is observed.

In the end, it should be noted that the comparison of the two downfolding
methods is at the level of the effective interactions. It will be very
interesting, in the future, to compare the ground state properties by
solving the effective one-band models, which requires advanced
numerical methods. We have presented here DMFT results which capture the
frequency dependence of the local density-density interactions. These results
showed that the effective correlation strength of the cRPA and cfRG downfolded
models is actually quite similar, despite the substantial differences in the static
values of $U$. This is the result of an opposite trend in the $\omega_n$-dependence
of the interactions. More advanced formalisms are needed to investigate the
effect of the non-density-density and nonlocal interactions induced by the downfolding.

Finally, we should mention that the exact downfolding not only produces a frequency-dependent
effective interaction, but also a renormalized bare propagator $G_0^\text{cRPA}$ or $G_0^\text{cfRG}$.
The bandwidth is usually reduced by self-energy corrections,\cite{Imada-2010,Hirayama-2018}
but the momentum dependence of the renormalization can be nontrivial. 
These renormalization effects are ignored in our paper. A reduction of the bandwidth would lead to stronger correlations in the
low-energy model, which could further enhance the discrepancy between the cfRG and cRPA downfolded models.
The investigation of this effect is left for a future study.

\begin{acknowledgments}
CH thanks DFG-RTG 1995 for support. 
PW acknowledges support from SNSF Grant No.~200021-165539.
We thank K. Held and G. Schober for discussions.

\end{acknowledgments}

\end{document}